\documentclass[%
preprint,
superscriptaddress,
 amsmath,amssymb,
 aps,
]{revtex4-1}

\usepackage{graphicx}
\usepackage{dcolumn}
\usepackage{bm}

\begin{document}


\title{Ion Clusters and Networks  in Water-in-Salt Electrolytes}

\author{Michael McEldrew}
\affiliation{Department of Chemical Engineering, Massachusetts Institute of Technology, Cambridge, MA, USA}

\author{Zachary A. H. Goodwin}
\affiliation{Department of Physics, CDT Theory and Simulation of Materials, Imperial College of London, South Kensington Campus, London SW7 2AZ, UK}
\affiliation{Department of Chemistry, Imperial College of London, South Kensington Campus, London SW7 2AZ, UK}

\author{Sheng Bi}
 \affiliation{ State Key Laboratory of Coal Combustion, School of Energy and Power Engineering, Huazhong University of Science and Technology (HUST), Wuhan, Hubei 430074, China}

\author{Alexei A. Kornyshev}
\affiliation{Department of Chemistry, Imperial College of London, Molecular Sciences Research Hub, White City Campus, Wood Lane, London W12 0BZ, UK}
\affiliation{Thomas Young Centre for Theory and Simulation of Materials, Imperial College of London, South Kensington Campus, London SW7 2AZ, UK}
\affiliation{Institute of Molecular Science and Engineering, Imperial College of London, South Kensington Campus, London SW7 2AZ, UK}

\author{Martin Z. Bazant}
\email{bazant@mit.edu}
\affiliation{Department of Chemical Engineering, Massachusetts Institute of Technology, Cambridge, MA, USA}
\affiliation{Department of Mathematics, Massachusetts Institute of Technology, Cambridge, MA, USA}

\date{}

\begin{abstract}
Water-in-salt electrolytes (WiSEs) are a class of super-concentrated electrolytes that have shown much promise in replacing organic electrolytes in lithium-ion batteries. At the extremely high salt concentrations of WiSEs, ionic association is more complicated than the simple ion pair description. In fact, large branched clusters can be present in WiSEs, and past a critical salt concentration, an infinite percolating ionic network can form spontaneously. In this work, we simplify our recently developed thermodynamic model of reversible ionic aggregation and gelation, tailoring it specifically for WiSEs. Our simplified theory only has a handful of parameters, all of which can be readily determined from simulations. Our model is able to quantitatively reproduce the populations of ionic clusters of different sizes as a function of salt concentration, the critical salt concentration for ionic gelation, and the fraction of ions incorporated into the ionic gel, as observed from molecular simulations of three different lithium-based WiSEs. The extent of ionic association and gelation greatly affects the effective ionic strength of solution, the coordination environment of active cations that is known to govern the chemistry of the solid-electrolyte interface, and the thermodynamic activity of all species in the electrolyte.
\end{abstract}

\maketitle

\section{Introduction}
Recently developed Water-in-Salt Electrolytes (WiSEs) have received a great deal of attention due to their potential for application in lithium-ion batteries (LIBs)~\cite{Suo2013,Sodeyama2014,Suo2015,Yamada2016,Wang2016,Gambou-Bosca2016,Sun2017,suo2017water,Dong2017,Diederichsen2017,yang2017a,Yang2017,wang2018hybrid,leonard2018,yamada2019advances,Yang2019,Dou2019,lewis2020signatures}. WiSEs are super-concentrated electrolytes often containing a mere 2-3 water molecules per ion pair. Typically, WiSEs contain small, alkali cations (such as Li$^+$) and bulky, fluorinated anions [such as TFSI$^-$, Bis(trifluoromethane)sulfonimide]. Owing to their high concentration, WiSEs have displayed enhanced electrochemical stability windows (ESW) compared to their dilute counterparts, making them promising candidates to replace the flammable organic electrolytes that are widely used in LIBs. Additionally, WiSEs tend to have superior transport properties compared to organic LIB electrolytes~\cite{borodin2017liquid}. We direct the reader to refs. \citenum{borodin2020uncharted,li2020new,chen2020water} for reviews highlighting recent progress in the literature on WiSEs.

There have been a number of computational studies highlighting the interfacial~\cite{vatamanu2017,mceldrew2018,li2018} and bulk~\cite{borodin2017liquid,choi2018,lim2018,yu2020asymmetric,lewis2020signatures,andersson2020ion} behavior of WiSEs, which have given valuable insight into the physicochemical and electrochemical behavior of WiSEs. However, WiSEs have not received enough theoretical attention in the literature. From a theoretical point of view, WiSE's embody the middle ground between solvent-less ionic liquids (ILs) and traditional solvent-dominated electrolytes. Theories of traditional electrolytes are well-established and have largely been rooted in the classical theories of Debye and H\"uckel~\cite{debye1923theory,huckel1925theorie}. In the past decade, ILs have received increasing attention from a theoretical point of view as highly correlated ionic matter as well as variety of their applications.~\cite{Fedorov2014,DongRev2017}. However, theories of traditional electrolytes and ILs often require simplified pictures of ion solvation and ionic aggregation. For example, in traditional electrolytes the solvent is often modeled as a dielectric continuum that is dependent on salt concentration and/or electrostatic fields~\cite{huckel1925theorie,fawcett1996role,abbas2007restricted,vincze2010nonmonotonic,shilov2015role}, and ion association has been largely limited to the formation of ion pairs~\cite{bjerrum1926k,ebeling1980analytical,levin1996criticality,marcus2006ion}. Prior to Ref.~\citenum{mceldrew2020corr}, ionic aggregation in ILs was treated only through the remaining amount of ``free" ions not in ionic aggregates~\cite{Chen2017,feng2019free,goodwin2017underscreening,Goodwin2017a} or through the formation of ion pairs~\cite{Ma2015,lee2014room}, and in rare cases small clusters of ion pairs~\cite{avni2020charge}. For WiSE's, the solvent is neither abundant nor negligible, but rather it is a discrete yet essential component of the solution that should be considered on the same  footing as ions in the electrolyte. Similarly, the coordinating environment an ion is neither a pristine solvation shell of solvent (dilute electrolyte) nor counter-ion (ionic liquid/crystal), but rather a complex combination of counter-ion and solvent coordination  \cite{pivnic2019structural}. Moreover, ions can be clustered in complex, branched aggregates that may even percolate throughout the mixture~\cite{choi2017ion,borodin2017liquid,choi2018,lim2018,lewis2020signatures,mceldrew2020theory}.
 

In this work, we apply our recently developed theory of ionic aggregation and gelation of super-concentrate electrolytes~\cite{mceldrew2020theory} to model WiSEs. The model accounts for the competition between ion association and ion solvation, finite molecular volumes, and the free energy to form ionic associations. We use our model to analyze ion aggregation, solvation, and gelation in various lithium-based aqueous electrolytes for salt concentrations ranging from the dilute regime to the water-in-salt regime. Using molecular dynamics (MD) simulations, we calculate the extent of ionic aggregation and solvation, distribution of ionic cluster sizes, and even the presence and extent of percolating ion networks. Our theory only depends on a handful of physically transparent parameters, such as the volume of ions and the number of ``bonds" that they can form and their strength. The performed MD simulations were also used to calculate these parameters. Overall, we find excellent agreement between our theory (with parameters obtained from MD) and the molecular simulations, which gives one confidence in the predictive power of the theory.


\section{Theory}
In our previous works, we developed a general theory of ion aggregation in super-concentrated electrolytes~\cite{mceldrew2020theory}, and a particular application of which was to ILs~\cite{mceldrew2020corr}. The general theory~\cite{mceldrew2020theory} accounted for associations between ions, forming complex and potentially even percolating ion networks, as well as the hydration of ions by solvent molecules, with bonds between solvent molecules being neglected.

Here, we simplify that general theory~\cite{mceldrew2020theory} by making several assumptions, which will be largely valid for super-concentrated WiSEs: 
\begin{enumerate}
    \item Selective bonding of water. We neglect anion solvation, as the anions used in WiSE's are bulky and relatively hydrophobic. Since the bonds typically form to the oxygens of the anions (as shown later), a consistent approximation is to also neglect bonding between water molecules. The water is only able to bind to the lithium cations, which means that secondary hydration shells are not accounted for. While this is certainly a simplified picture of the solution structure, we believe the Li$^+$-OH$_2$ coordination to be the dominant effect.
    \item Simplified electrostatics and interactions. We neglect the excess electrostatic energy of the mixture, such as the Debye-H\"uckle term of dilute electrolytes~\cite{debye1923theory,huckel1925theorie}. This assumption relies on the electrostatic energy being dominated by the formation of ionic clusters, which are assumed to behave ideally in the solution (no inter-cluster enthalpic interactions). This simplification was also employed successfully in Ref.~\citenum{mceldrew2020corr} to model ILs, and we expect it to work reasonably well for super-concentrated WiSEs. Moreover, in Ref.~\citenum{mceldrew2020theory}, the effect of Debye-H\"uckle screening and Born solvation of free ions was studied and was found to not significantly affect ion cluster distributions at high concentrations. 
    \item``Sticky" cation approximation. Our last simplification is that the cations always have a full coordination shell, comprising of water molecules and/or anions. As we will show, this assumption is especially valid for lithium-based WiSEs, in which lithium cations interact very strongly with both the water and the anion. 
\end{enumerate}
\noindent For brevity, we will simply outline the crux of theory in the main text, and relegate some of the details to the supporting information (SI).

We consider a polydisperse mixture of $\sum_{lms}N_{lms}$ ionic clusters containing $l$ cations, $m$ anions, and $s$ water molecules (a rank $lms$ cluster), and an interpenetrating gel network (if present) containing $N_+^{gel}$ cations, $N_-^{gel}$ anions, and $N_0^{gel}$ water molecules. As in Refs.~\citenum{mceldrew2020theory}~\&~\citenum{mceldrew2020corr}, we use a Flory-like lattice fluid free energy, which has been used extensively to describe associative polymer mixtures~\cite{tanaka1989,tanaka1990thermodynamic,tanaka1994,tanaka1995,ishida1997,tanaka1998,tanaka1999,tanaka2002}:
\begin{align}
    \beta \Delta F &= \sum_{l,m,s} \left[N_{lms}\log \left( \phi_{lms} \right)+N_{lms}\Delta_{lms}\right] \nonumber \\
    &+ \Delta^{gel}_+ N^{gel}_+ + \Delta^{gel}_- N^{gel}_- + \Delta^{gel}_0 N^{gel}_0 
    \label{eq:F}
\end{align}
where $\beta=1/k_BT$ is inverse thermal energy, $\phi_{lms}$ is the volume fraction of rank $lms$ clusters, $N_{lms}$ is the number of rank $lms$ clusters, $\Delta_{lms}$ is the free energy of formation for clusters of rank $lms$, and $\Delta^{gel}_i$ is the free energy change of species $i$ upon association to the gel. The free energy in Eq.~\eqref{eq:F} contains just three essential pieces of physics: the ideal entropy of mixing for the distribution of clusters (first term), the free energy of formation of the clusters (second term), and the free required for species to join the gel (last three terms).

The WiSE is modelled to reside on a lattice containing $\Omega$ lattice sites, with volume, $v_0$ (equivalent to the molecular volume of a water molecule). Anions and cations each require $\xi_+=v_+/v_0$ and $\xi_-=v_-/v_0$ lattice sites per cation or anion, respectively, where $v_i$ is the molecular volume of species $i$. Thus, the volume fraction of a rank $lms$ cluster can be written simply as $\phi_{lms}=(\xi_+l+\xi_-m+s)N_{lms}/\Omega$. 

The ions and water molecules are modelled to form associations via a fixed number of association sites, known as the species functionality, $f_i$. We model the cations to have a functionality of $f_+$, and anions to have a functionality of $f_-$. We assume that water molecules coordinate to cations with a functionality of 1 ($f_0=1$). The clusters that are formed are assumed to be Cayley trees (branched with no intra-cluster loops), which are decorated with water molecules residing on the open cationic association sites. Thus, the backbone of the cluster purely comprises of ions. Any cluster containing $l$ cations and $m$ anions will contain exactly $l+m-1$ ionic associations. As we assume the cations here to be ``sticky", the number of water molecules in a cluster, $s$, is thus the total number of cationic association sites in the cluster ($f_+l$) minus the number of ionic associations ($l+m-1$), i.e., $s=f_+ l -l -m +1$. Note that because $s=f_+l-l-m+1$, the rank of the cluster, $lms$, can actually be specified by just the ion indices, $lm$. However, for book-keeping reasons, we will continue to use $lms$ to specify the cluster index. 
The equilibrium distribution of clusters can be derived by enforcing chemical equilibria (shown explicitly in the SI), yielding the following relation:
\begin{align}
    \phi_{lms}=K_{lms}\phi_{100}^l \phi_{010}^m \phi_{001}^s
    \label{eq:cluster_dist}
\end{align}
where $K_{lms}=\exp(l+m+s-1-\Delta_{lms})$ is the cluster equilibrium constant. Thus, the equilibrium cluster distribution depends on the explicit form of $\Delta_{lms}$, which is derived in the SI. The strength of the presented theory comes from the fact that $\Delta_{lms}$ only requires six, physically transparent parameters [$\xi_{\pm}$, $f_{\pm}$, $\tilde{\lambda}$ (the ionic association constant) and $\tilde{c}_{salt}$ (dimensionless concentration of salt - $\#$ per lattice site)] to be specified to determine all possible $\Delta_{lms}$. 

Provided the functionalities, $f_\pm$, of anions and cations are greater than two, the clusters can become infinitely large. For this to occur, the probability that a cation (anion) has another cation (anion) separated by an anion (cation) must reach 1. The point at which this occurs (i.e. the gelation point) is given by the condition
\begin{align}
p^*_{+-}p^*_{-+}(f_+-1)(f_--1)=1
\label{eq:gelpt}
\end{align}
where $p^*_{+-}$ and $p^*_{-+}$ are the critical (denoted with $*$) ion association probabilities for gelation. The ion association probability, $p_{ij}$ is defined as the fraction of association sites on species $i$ occupied by an association to species $j$. Equation~\eqref{eq:gelpt} is a classical result of Flory~\cite{flory1941molecular} and Stockmayer~\cite{stockmayer1943theory} and is analogous to bond percolation on a Bethe lattice which arises when the cumulative probability of a cluster to branch indefinitely becomes unity. These association probabilities essentially close the theory, and make the connection between microscopic bonding picture and bulk thermodynamics. As shown in the SI, these association probabilities can be explicitly expressed as a function of salt concentration:
\begin{align}
    f_+p_{+-}=f_-p_{-+}=\frac{1 + \mathcal{D}\tilde{c}_{salt} -\sqrt{(1+\mathcal{D}\tilde{c}_{salt})^2+4f_-\tilde{\lambda}\tilde{c}_{salt}}}{2\tilde{c}_{salt}(\Tilde{\lambda}-1)} 
\end{align}
where $\mathcal{D}= f_+(\tilde{\lambda}-1)+ f_-\tilde{\lambda} - \xi_+ - \xi_-$
and the ionic association constant 
\begin{align}
\tilde{\lambda}=\exp\left\{-\beta(\Delta U_{+-}-\Delta U_{+0})\right\}
\end{align}
governs the extent of ion association, where $\Delta U_{+-}$ is the energy of a cation-anion association and $\Delta U_{+0}$ is the energy of a cation-water association. The ion association constant is determined by the difference between of $\Delta U_{+-}$ and $\Delta U_{+0}$ as a direct result of the ``sticky" cation approximation: every time a cation-anion association is formed, a cation-water association must be destroyed. The cation-anion and cation-water associations are largely electrostatic owing to the monopolar and dipolar nature of the ions and water molecules, respectively. 

Also as a consequence of the ``sticky" cation approximation, the probabilities corresponding to cation-water associations, $p_{+0}$ and $p_{0+}$, can be directly related to the ion association probabilities:
\begin{align}
    p_{+0}=\left(\frac{\tilde{c}_{water}}{f_+\tilde{c}_{salt}}\right)p_{0+}=1-p_{+-}
\end{align}
where the dimensionless concentration of water is given by $\tilde{c}_{water}=1-(\xi_++\xi_-)\tilde{c}_{salt}$, as prescribed by our assumption of electrolyte incompressibility.

In WiSEs, the salt concentration is readily altered through the addition/removal of salt or water. At low salt concentrations, one does not expect to form an ionic gel, but in highly concentrated WiSEs that are employed in battery applications, a gel could potentially form. From solving Eq.~\eqref{eq:gelpt}, we can obtain the critical salt concentration, $\tilde{c}^*_{salt}$, required to form an ionic gel:
\begin{align}
    \tilde{c}^*_{salt}=(\mathcal{A}\tilde{\lambda}+\mathcal{B})^{-1}
\label{eq:crit_salt}
\end{align}
where $\mathcal{A}$ and $\mathcal{B}$ are constants given by
\begin{align}
    \mathcal{A}=\frac{1}{2}\sqrt{\frac{
  f_+ f_-}{(f_+ - 1)(f_- - 1)}} + 2 \sqrt{f_+ f_- (f_+ - 1) (f_- - 1)} - f_+ - f_-
\end{align}
and 
\begin{align}
    \mathcal{B}=\xi_+ + \xi_- - \frac{1}{2} \sqrt{\frac{f_+ f_-}{(f_+ - 1) (f_- - 1)}}
\end{align}
Alternatively, one can think about fixing the concentration of salt in the WiSE and varying the temperature. One can invert Eq.~\ref{eq:crit_salt} and obtain a critical value of $\tilde{\lambda}^*=-\mathcal{B}+1/\mathcal{A}\tilde{c}_{salt}$, which would directly correspond to a critical gel-point temperature for a WiSE at a specific salt concentration:
\begin{align}
    T*=\left(\frac{\Delta U_{+-}-\Delta U_{+0}}{k_B}\right)(\mathcal{B}+1/\mathcal{A}\tilde{c}_{salt})^{-1}
\end{align}
This equation can be used to predict the critical gel line in the concentration--temperature plane. However, we must be sure to take extreme care to ensure that the  sticky cation approximation is applicable across the studied temperature range. At high temperatures, it becomes more likely that association sites on the cation are unoccupied (with either anion or water), which could have a considerable effect on the predicted critical gel line. 

When the gel is present, it becomes necessary to determine the exact fraction of species in the gel, as well as the distribution of finite clusters not incorporated in the gel (i.e. sol). This computation requires an slightly altered procedure than what was outlined in this section. For the details outlining this procedure, as well as an extended discussion of the critical gel line in the concentration--temperature plane, we direct the reader to the SI.

\section{Results}
The theory presented in the preceding section allows us to predict the ion clustering and solvation in WiSEs. In order to test the validity of our developed theory, we performed a series of molecular dynamics (MD) simulations of aqueous lithium bis(trifluoromethanesulfonyl)imide (LiTFSI), lithium triflouromethanesulfonate (LiOTF), and lithium bis(fluorosulfonyl)imide (LiFSI) electrolytes for concentrations ranging from the dilute to ``water-in-salt" regime. In this section, we summarize our results and observations from these MD simulations, and then we compare those results against our theory. We mainly focus on the results of LiTFSI in the main text, to avoid repeating similar observations in the other salts. The results for LiOTF and LiFSI solutions are mainly shown in the SI.


\subsection{Criteria and Characteristics of Association}
\begin{figure}[hbt!]
    \centering
    \includegraphics[width=\textwidth]{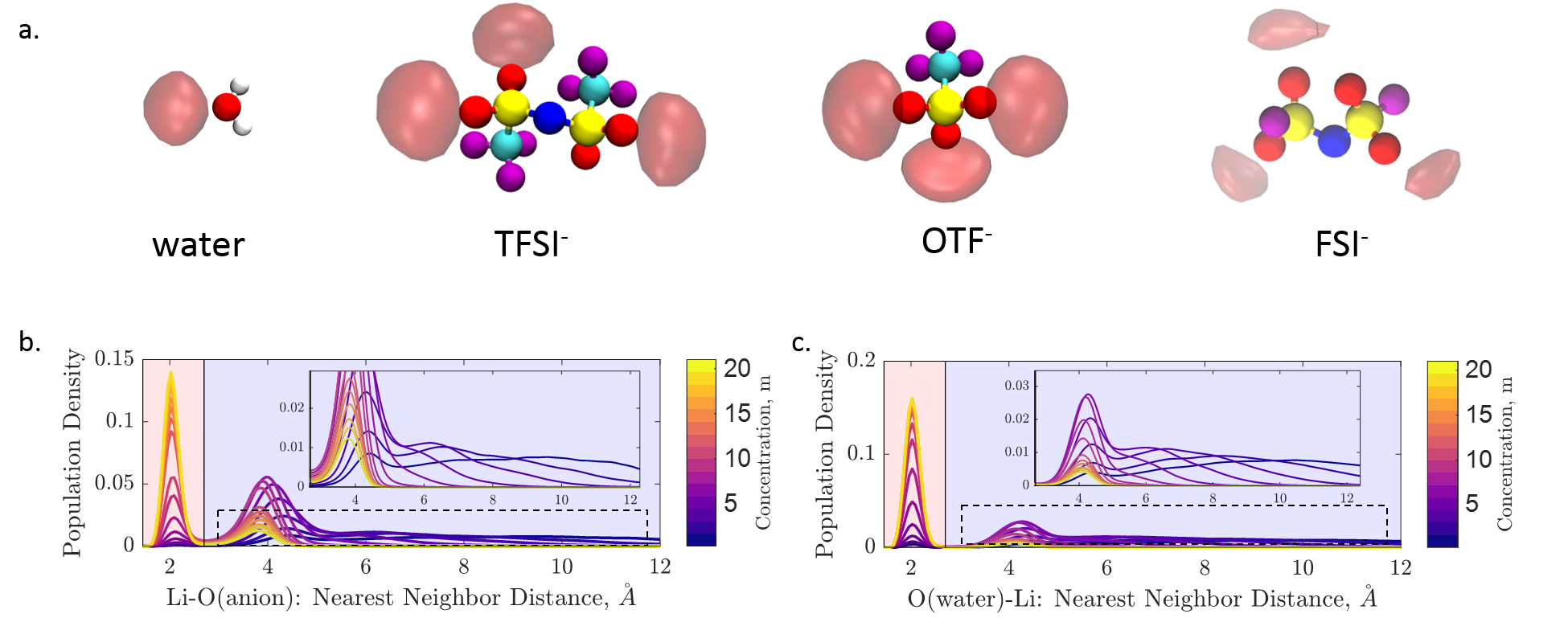}
    \caption{(a) The spatial distribution functions of Li$^+$ around water, TFSI$^-$, OTF$^-$, and FSI$^-$ showing iso-density surfaces corresponding to 4x the bulk density of Li$^+$ .
    The probability distribution of the nearest-neighbor distances for (b) Li$^+$--O(TFSI$^-$) and (c) O(water)--Li$^+$. The insets in (b) and (c) zoom in on the probability distributions within the boxed region. }
    \label{fig:rnn}
\end{figure}

In order to study the degree of ionic aggregation and solvation in these electrolytes we first need a precise definition of what constitutes a lithium-water association and a lithium-anion association. The Li$^+$ interacts very strongly with the oxygen of the water molecule, as well as the oxygens of the TFSI$^-$, OTF$^-$, and FSI$^-$ ions. The affinity of Li$^+$ to water or anionic oxygens can be seen clearly by the spatial distribution functions visualized in Fig.~\ref{fig:rnn}a as iso-density surfaces. These iso-density surfaces encapsulate regions around the water molecule or anion where Li$+$ density is on average \emph{at least} 4$\times$ its bulk density. It can be seen clearly that these regions are typically around the oxygens of the TFSI$^-$, OTF$^-$, and FSI$^-$ ions. Thus, it is natural to define a lithium-anion or lithium-water association based on the proximity of Li$^+$ to oxygen atoms, belonging either to anions or water, respectively. In particular, we use a Li$^+$-O distance threshold of 2.7 $\AA$ to determine if a Li$^+$ is associated to oxygens on water molecules or anions. 

This distance threshold is not arbitrary, and can be understood clearly when examining the nearest neighbor distance distributions between the associating species in the electrolyte. Here we will define an A--B nearest neighbor distance distribution to be the probability of finding a species A, whose nearest neighbor, B, is a distance, r, away from A.  In Fig.~\ref{fig:rnn}b~\&~c, we plot the Li$^+$--O(TFSI$^-$) (Fig.~\ref{fig:rnn}b) and the O(water)--Li$^+$ (Fig.~\ref{fig:rnn}c) nearest neighbor distance distributions for concentrations of LiTFSI ranging from  0.28m--21m.

For the Li$^+$--O(TFSI$^-$) nearest neighbor distance distribution first has a sharp peak at a distance of roughly 2~$\AA$, approximately corresponding to the distance of closest approach between lithium and oxygen. The distribution then decreases to a minimum at 2.7~$\AA$, before we observe another peak centered around distances of roughly 4~$\AA$, which becomes sharper as the concentration increases. Thus, there exists one sub-population of lithium in direct contact with anion and another sup-population of lithium not in direct contact with anion. We have a near identical story for the O(water)--Li$^+$nearest neighbor distance distribution. The distribution is indeed bi-modal, with two peaks again corresponding to sub-populations of water molecules that are in direct contact (2~$\AA$) with lithium and another that is not in direct contact ($\sim$4~$\AA$) with lithium, separated by probability minimum at roughly 2.7~$\AA$. Thus, our Li$^+$-O cutoff distance of 2.7~$\AA$ provides a clear-cut criterion for both Li$^+$-anion and Li$^+$-water associations.


The insets of Fig. \ref{fig:rnn}~b\&c zoom in on the behavior of the nearest neighbor distributions at large distances. We observe that at high concentrations ($>$2~m), the 
Li$^+$--O(anion) nearest neighbour distance distributions rapidly decay to zero for nearest-neighbor distances greater than 4$\AA$. Therefore, at these high concentrations, it becomes exceedingly likely that lithium is present only in one of two states: in direct contact with an anionic oxygen (2~$\AA$), and indirect second-shell contact with an anionic oxygen (4~$\AA$). However, at low salt concentrations ($<$2~m), the Li$^+$--O(anion) nearest-neighbor distance distributions do not rapidly decay after the second maximum at 4~$\AA$. Instead, they actually observe a third peak in the nearest neighbor distribution, which is much more spread out. Thus, it is clear that at concentrations less than 2~m there are 3 characteristic states in which ions exist: direct first-shell contact (2~$\AA$), indirect second-shell (i.e. solvent-separated ~\cite{eigen1962schallabsorption,marcus2006ion,seo2012electrolyte1,seo2012electrolyte2}) contact (4~$\AA$), and free ($>$6~$\AA$). In all cases, the probability for any lithium ion to have its nearest anionic oxygen be very large distances ($>12\AA$) away becomes extremely small. 

We should note that there is no distinction in our theoretical model between ions that are in indirect second-shell contact (4~$\AA$) and completely free ions($>$6~$\AA$); these scenarios would both correspond to ions that are not considered to be associated. The exchange between second-shell and free states could likely be modelled well within Bjerrum's framework of ion pairing~\cite{bjerrum1926k}, but it would only be useful to model this exchange at low salt concentrations. For concentrations greater than 2~m, the probability of any ions being significantly separated ($>$6~$\AA$) becomes exceedingly rare, and the majority of ions exist only in direct first-shell contact or indirect second-shell contact. For these high salt concentrations, our binary definition of ion association becomes much more appropriate; nearest ionic neighbors are either in the first shell or the second shell.

In Fig.~\ref{fig:rnn}e, there is a similar story for the  O(water)--Li$^+$ nearest-neighbor distance distributions.  At all concentrations there are peaks corresponding to water molecules in the first (2~$\AA$) and second solvation (4~$\AA$) shells of Li$^+$. However, for concentrations less than 2~m, (as seen in the inset of Fig. \ref{fig:rnn}~c), we observe a peak in the O(water)--Li$^+$ nearest neighbor distribution corresponding to water molecules not within the the first two coordination shells of any Li$^+$, i.e. truly free water. Again our model treats water molecules outside of the first coordination shell as not being associated to Li$^+$ and does not distinguish between second shell water molecules and truly free water molecules (outside of the second solvation shell of all Li$^+$). There are existing statistical models based on multi-layer Langmuir adsorption~\cite{dutcher2011statistical} that can handle multi-shell ion hydration. However, such an extension to our model would only be necessary for low salt concentrations, as it becomes exceedingly unlikely for water molecules to exist outside the second shell of Li$^+$ for concentrations greater than 2~m. 

Thus, our simulations show that our model's simple binary picture of species as associated or not associated (either to ion or water) is not strictly true for low salt concentrations. However, this binary picture does become valid at high salt concentrations, in which species reside almost strictly two distinct states corresponding to species that are associated or unassociated. The primary focus of this paper is on super-concentrated WiSE's, where our model is likely to be largely valid, as indicated by our simulations.  




\subsection{Li$^+$ Coordination: Ionic Association vs. Solvation}
\begin{figure*}[hbt!]
    \centering
    \includegraphics[width=1\textwidth]{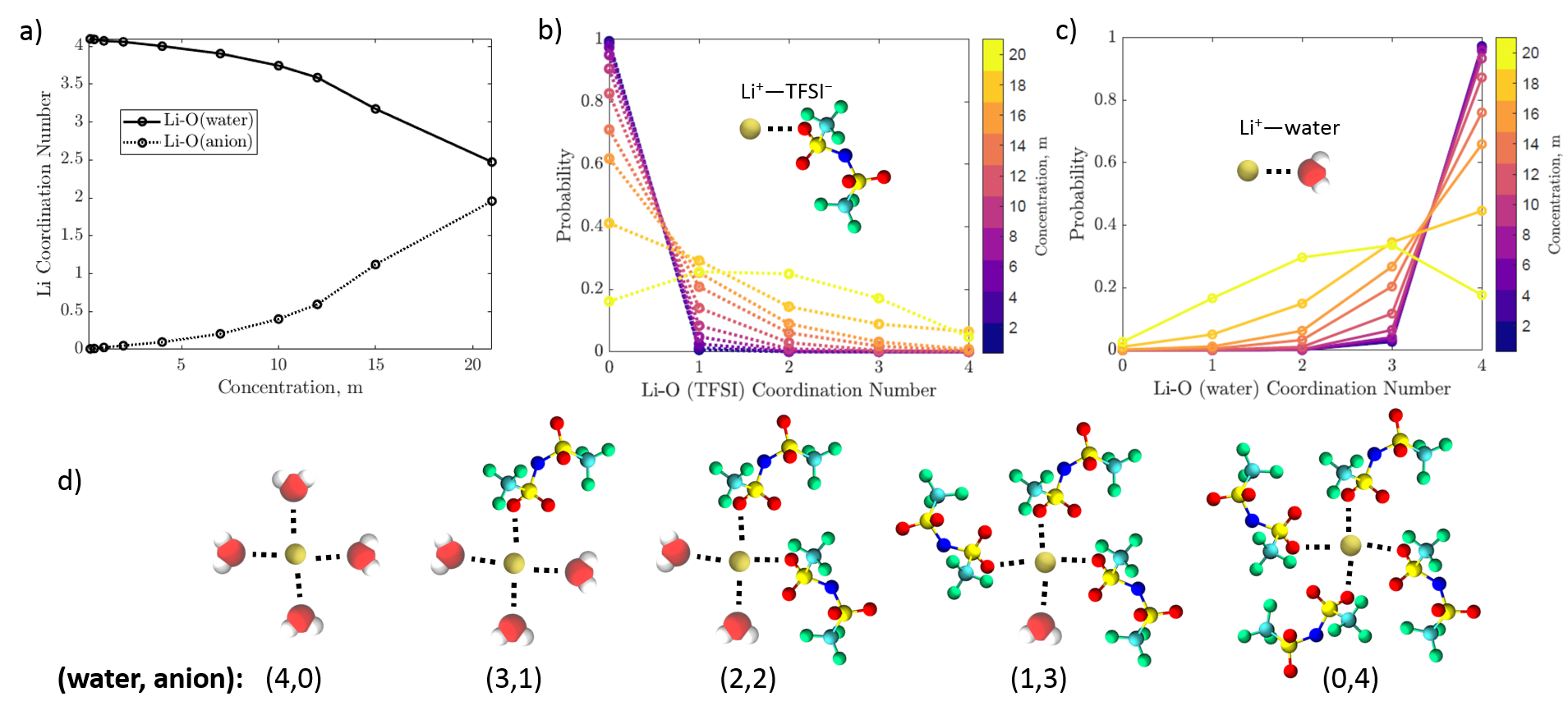}
    \caption{(a) The average coordination number of Li$^+$by water oxygens and by TFSI$^-$ oxygens as a function of concentration, showing the competitive nature of ion association and solvation. (b) The probability distributions for the number of TFSI$^-$ oxygens coordinating lithium. (c) The probability distributions for the number of water oxygens coordinating lithium. Distributions are plotted for LiTFSI concentrations of 0.28~m-21~m. (d) A schematic showing cartoon depictions of typical coordination environments of Li$^+$ in which Li$^+$ is coordinated by 4 oxygen atoms either from TFSI$^-$ or water. }
    \label{fig:licoord}
\end{figure*}

A very important phenomenon in WiSEs is the competition between ionic association and solvation. From our simulations, we find that Li$^+$ ions generally have four associations to oxygen atoms; either to the anions or water molecules, resulting in a competition between Li$^+$--O(anion) and Li$^+$--OH$_2$ coordination due to the limited space around the Li$^+$ ions. If Li$^+$--OH$_2$ associations dominate over Li$^+$--O(TFSI$^-$), then ion aggregation is suppressed and the formation of an ionic gel is forestalled. The competitive coordination can be seen clearly in Fig.~\ref{fig:licoord}a, where we have measured (via MD simulation) the average coordination number of Li$^+$ to the water oxygen or TFSI$^-$ oxygens as a function of the salt concentration. We observe that the average Li$^+$--OH$_2$ and Li$^+$--TFSI$^-$ coordination numbers respectively decrease and increase monotonically as a function of salt concentration. This is further illustrated in Fig.~\ref{fig:licoord}b~\&~c, where the Li$^+$ coordination number probability distributions are plotted for both water oxygens (Fig.~\ref{fig:licoord}b) and TFSI$^-$ oxygens (Fig.~\ref{fig:licoord}c). For any given concentration, the probability distributions for Li$^+$ coordination by anion and water have a strong reflective symmetry with each other, which is a result of the approximately conserved Li$^+$-O coordination number of four. Note, that this Li$^+$-O coordination number conservation is in strong agreement with the ``sticky" cation approximation that was a main assumption of the derived theory.

This concept is further depicted in Fig.~\ref{fig:licoord}d, in which we draw the 5 possible coordination environments that conserve 4 associations to Li$^{+}$. Note that Li$^+$ coordination is not the complete picture. Each TFSI$^-$ ion coordinating Li$^+$ contains additional oxygen atoms that can associate with further Li$^+$. On the other hand, each of the hydrating water molecules serve as a network terminating agent because they can only associate to a single Li$^+$ ion. Thus, Li$^+$ coordination informs us about the average number and types of associations that Li$^+$ makes, but it does not tell us the extent and size of clusters that are formed. 

\subsection{Ionic Aggregation and Gelation}

\begin{figure*}[hbt!]
    \centering
    \includegraphics[width=1\textwidth]{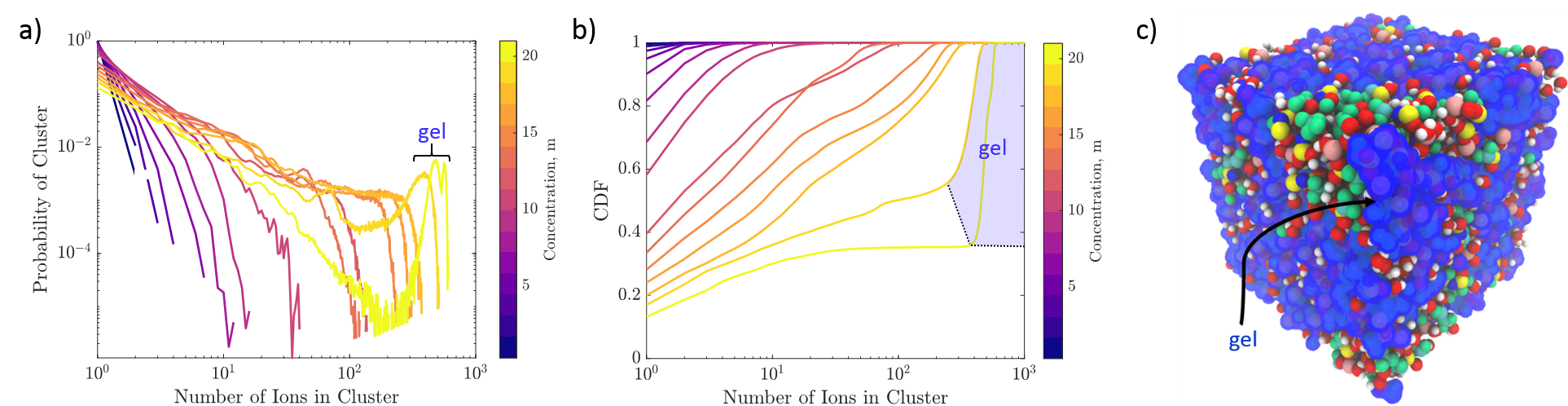}
    \caption{(a) The probability distribution of ionic cluster sizes for a range of different concentrations (0.2~m-21~m) of LiTFSI. Peaks in the distributions for concentrations greater than 16~m correspond to percolating ionic gels. (b) The cumulative distribution function (CDF) of ionic cluster sizes for the various concentrations. The shaded region corresponds to ionic gel which result in a sharp uptick in the CDF. (c) A visualization of an MD snapshot of 21~m LiTFSI. The ionic gel is highlighted in blue, and is observed to percolate throughout the entire simulation box.}
    \label{fig:ion_agg}
\end{figure*}

As was previously mentioned, the fact that Li$^+$ and TFSI$^-$ (as well as OTF$^-$ and FSI$^-$), are able to form multiple associations permits the formation of large, potentially infinite, ion networks. In Fig.~\ref{fig:ion_agg}a we plot the probability distribution, $\mathcal{P}_{n}$, for an ion to be incorporated in an ion cluster containing a total of $n$ ions (anions + cations), for a range of different LiTFSI concentrations. We see that for concentrations below 19~m, $\mathcal{P}_{n}$ decreases monotonically as a function of the cluster size. This observation might be intuitively expected as larger clusters require the coordination of a larger number of ions, which would seem increasingly unlikely at low concentrations. For increasing salt concentrations (less than 19~m), it becomes increasingly more likely to observe higher order clusters. This is intuitive because ions are more likely to associate when there is more salt and less water in the mixture. 

At concentrations above 19~m, the cluster size distributions are no longer monotonically decreasing. Rather, we observe a remarkable peak in the distribution corresponding to extremely large ionic clusters. This observation is illustrated more clearly by the cumulative distribution function (CDF) plotted in Fig.~\ref{fig:ion_agg}b, where we see a seemingly anomalous uptick in the CDF indicated by the dashed line enclosing the shaded region. In fact, the clusters within the probability distribution peaks (indicated by the gel label), as well as the shaded region of the CDF, correspond to clusters that involve a large fraction of the total amount of ions in the simulation. We can see this in Fig.~\ref{fig:ion_agg}c, where the largest cluster is highlighted in blue in a snapshot of a MD simulation of 21~m LiTFSI. The highlighted cluster extends throughout the entire simulation box, and we would expect it to grow arbitrarily large depending on the size of the simulation. Thus, we define this percolating structure as the \emph{ionic gel}. This graphic (in Fig.~\ref{fig:ion_agg}c) also illustrates that the ionic gel is far from crystalline. In fact, the gel is quite spatially disordered, which is characteristic of polymeric gels~\cite{flory1953principles}. This spatial disorder comes through the chaotopic anions, not the kosmotropic cations. Physically, what is propelling the percolation of the ion cluster is that ion associations become so abundant in the mixture that clusters have a finite probability to continue indefinitely as opposed to terminating. 



In polymeric mixtures, the presence of gel can lead to large changes in the mixture's physical properties, such as the divergence of viscosity and the onset of finite elasticity~\cite{flory1953principles}. In WiSE's the ionic gels are patched together by attractive electrostatic interactions, which are thermally reversible (breakable). Thus, we hypothesize that WiSE's containing gels are potentially viscoelastic fluids, perhaps able to bridge the gap between typical liquid electrolytes and polymeric solid-state electrolytes in a tunable fashion. However, the viscoelastic properties of WiSEs would depend highly on the lifetime and strength of the associations. For example, Winter \cite{winter2002critical} describe such physical gels as having solid-like behavior below a certain yield stress (related to the strength of associations) and at time scales shorter than the so-called renewal time of the gel (related to the mean association lifetime). We find that the association lifetimes are on the order of picoseconds (3-13ps for aqueous LiTFSI electrolyte as shown in figure S4 in the  SI). The reason for the relatively short association lifetimes is not because the associations are weak, but rather because there is a strong competition between ion association and lithium hydration, as we have explained, which tends to limit the lifetime of the ion associations. Actually, as shown in the subsequent section, the success of the sticky-cation approximation in describing the clustering in lithium-based WiSEs implies that ion associations are very strong, with magnitudes well above $k_BT$. Nonetheless, the picosecond scale lifetimes suggest that ion network-forming WiSEs could display solid like properties, but only below picosecond timescales. 

Perhaps of equal importance is what the association lifetime tells us about the ability of ion clusters to  contribute to diffusion in our system. In ref. \citenum{feng2019free}, it was shown that species contribute to diffusion only if their lifetimes are longer than the decay of their velocity  auto-correlation functions (VACFs). In ref. \citenum{mceldrew2020corr}, this idea was extended to ion clusters yielding an approximate constraint on the sizes of clusters that may contribute to the diffusive relaxation of the system:
\begin{equation}
    (l+m)^{2/3}(l+m-1)<\tau_B/\tau_\nu
    \label{eq:constraint}
\end{equation}
where $l$ is the number of cations in the cluster, $m$ is the number of anions in the cluster, $\tau_B$ is the association lifetime, and $\tau_\nu$ is the characteristic time of the decay of the VACF for free ions (generally around 1 ps). Thus, for ion association times of 3-13 ps (as is the case for aqueous LiTFSI electrolyte), we would expect only clusters containing 3-5 ions \emph{or less} could contribute diffusively to the system.

\subsection{Comparison of Theory and Simulation}
Our MD simulations demonstrate the relevance of ionic gelation, and competitive ionic association and solvation in WiSEs. Our thermodynamic model was designed to describe such phenomena. Thus, in this section, we test the ability of our model to capture our observations from the aforementioned simulations. 

Our model contains a number of molecular parameters: $\xi_\pm$, $f_\pm$, $\tilde{\lambda}$. The size parameters $\xi_\pm$ are obtained by computing the molecular volumes consistent with the force fields in our MD simulations (volume enveloped by the overlapping Lennard-Jones radii of the molecule). The Li$^+$ functionality was determined to be 4, which is consistent with the Li$^+$ coordination data in Fig.~\ref{fig:licoord}. Although we did observe rare instances of Li$^+$ coordinating more than 4 species (water+anionic oxygens), these instances were rare enough to still model Li$^+$ as having $f_+=4$. The anionic functionality can be reasoned clearly by the observed SDFs in Fig.~\ref{fig:rnn}a. Here we can see the regions around each of the anions with a specific preference for Li$^+$. For each anion, (TFSI$^-$, OTF$^-$, and FSI$^-$), we observed 3 distinct regions, indicating that they will have all have an ion functionality of $f_-=3$. This is slightly unexpected for TFSI$^-$ and FSI$^-$, because these anions have 4 oxygens that might associate to Li$^+$, which would ostensibly indicate that $f_-=4$ for these species. Nevertheless, it seems to be the case that only 3 of the anionic oxygens can bind an Li$^+$ at any one time. Again, there are rare instances where the anions associate to more than 3 Li$^+$ ions, but these instances are exceedingly rare. The functionality of the solvent, $f_0$ was determined to be 1.

\begin{figure}[hbt!]
    \centering
    \includegraphics[width=\textwidth]{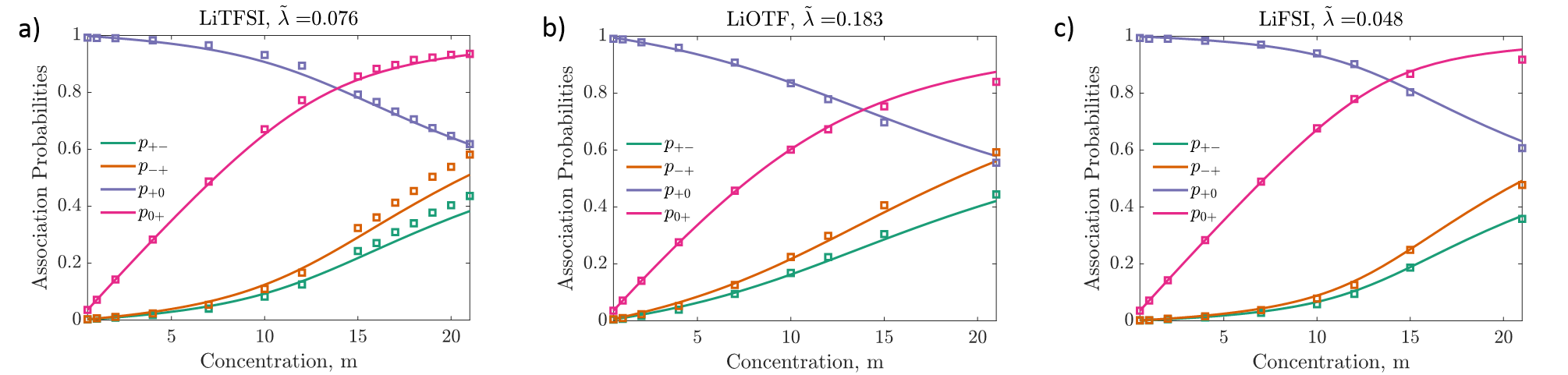}
    \caption{Association probabilities ($p_{+-}$, $p_{-+}$, $p_{+0}$, \& $p_{0+}$) are plotted as a function of salt concentration for (a) LiTFSI (b) LiOTF and (c) LiFSI electrolytes. The square data points are obtained from molecular dynamics (MD) simulations, and the solid curves are generated from our theoretical model, using association constants, $\tilde{\lambda}$, that are computed from the MD simulations. All curves were generated using the parameters, $\xi_+=0.4$, $f_+=4$, and $f_-=3$. LiTFSI curves used the additional parameters $\lambda=0.075$ and $\xi_-=10.8$. LiOTF curves used the additional parameters $\lambda=0.184$ and $\xi_-=5.9$. LiFSI curves used the additional parameters $\lambda=0.049$ and $\xi_-=6.8$.}
    \label{fig:association_consts}
\end{figure}

The remaining parameter, $\tilde{\lambda}$ can also be easily computed from MD simulation data. As is shown in the SI, the ion association constant is related to the association probabilities, $p_{ij}$, through a mass action law which describes the exchange of water and anions in the lithium coordination shell:
\begin{align}
    \tilde{\lambda}=\frac{p_{-+}(1-p_{0+})}{p_{0+}(1-p_{+-})}
    \label{eq:massac}
\end{align}
Thus, in order to compute $\tilde{\lambda}$ we need only compute the association probabilities from simulation. The association probabilities can be computes simply using the following formulae:
\begin{align}
    p_{ij}=\left\langle\frac{\# \text{ of associations of type }ij }{f_i \cdot \# \text{ of molecules of type }i}\right\rangle
\end{align}
where the $\langle \cdot \rangle$ operation indicates an ensemble average. The probabilities are obtained as a function of salt concentration, and then $\tilde{\lambda}$ is computed from Eq.~\eqref{eq:massac} and averaged across all concentrations. 

In Fig.~\ref{fig:association_consts}, we plot the association probabilities, $p_{ij}$, as a function of salt concentration for each of the simulated electrolytes, LiTFSI, LiOTF, and LiFSI. For each salt, the theory curves (with model parameters computed from MD) match the simulated values almost quantitatively in the entire concentration range studied. 

\begin{figure*}[hbt!]
    \centering
    \includegraphics[width=\textwidth]{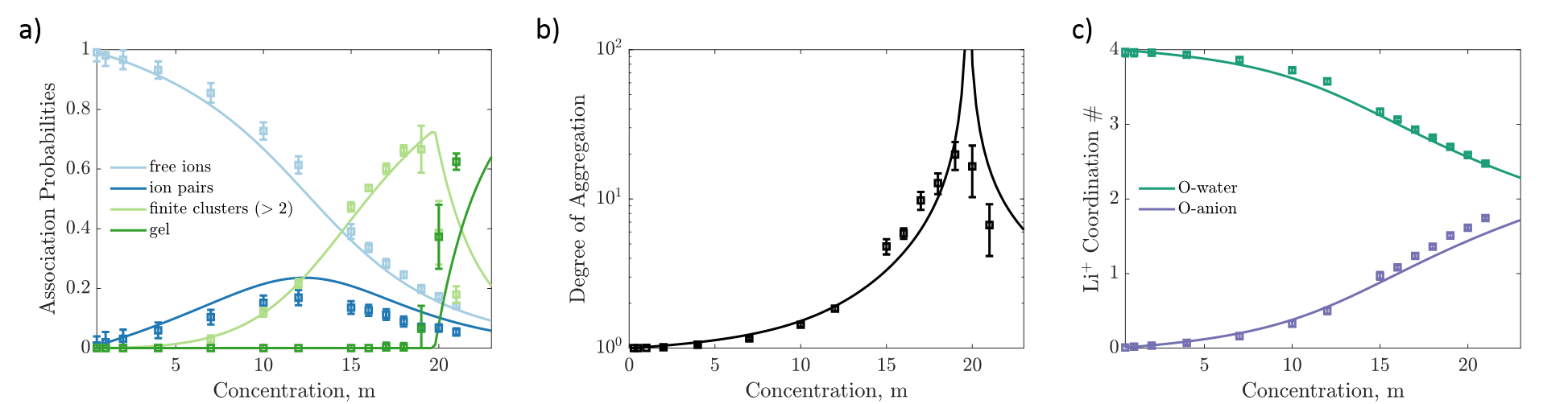}
    \caption{(a) The fractions of various ionic cluster types (free ions, ion pairs, finite high order clusters and ionic gel), (b) the weight-average degree of aggregation (for finite clusters), and (c) the coordination numbers of Li$^+$ by oxygen atoms from water and oxygen atoms by TFSI$^-$,. The curves were generated using the parameters $\lambda=0.075$, $\xi_+=0.4$, $\xi_-=11.3$, $f_+=4$, and $f_-=3$.}
    \label{fig:comp}
\end{figure*}

With the parameters in our model fully specified, we may explore theoretical predictions of the model and test their accuracy with MD. Of foremost importance, is the clustering of ions as a function of salt concentration. In Fig.~\ref{fig:comp}a, we plot the concentration dependence of the fractions of various ion cluster types: free ions (not bound to any other ions), ion pairs (one cation bound to one anion), finite clusters (clusters larger than ion pairs up to, but excluding, the gel) and ionic gel (as previously defined). The model does a remarkable job reproducing the trends we observe from simulations for all of the plotted ion cluster types.

As expected for aqueous electrolytes in dilute to moderate salt concentrations ($<$5~m), the electrolyte is dominated by free ions, and the ion aggregates that appear are almost entirely ion pairs. In this regime, ion aggregation would be consistent with Bjerrum's description of ion pairing. However, upon increasing the salt concentration to $\sim$10~m, we observe that ions are more likely to appear in high order clusters than simple ion pairs. Though, the free ions still comprise the majority of the total ions in the electrolyte. Further, increasing the concentration up to roughly 19~m, we find that ions are more likely to be present in high order finite clusters than both ion pairs and free ions. In fact, the fraction of ions appearing in ion pairs begins to decay, with the max fraction of ion pairs occurring at a concentration of roughly 12~m. Finally, for concentrations greater than roughly 19~m, we enter the ionic gel regime. The gel point is characterized in Fig.~\ref{fig:comp}a not only by the onset of a finite gel fraction, but also by a sharp decrease in the fraction of ionic cluster larger than ion pairs, suggesting that large portions of the high order finite aggregates combine to form the gel at and beyond the gel point.


The weight-averaged degree of ionic aggregation, $\bar{n}_w$, which is plotted as a function of salt concentration in Fig.~\ref{fig:comp}b, also provides insight into the extent of clustering. The weight average degree of aggregation is defined as the following
\begin{align}
    \bar{n}_w=\langle l+m\rangle=\frac{\sum_{lm}(l+m)\alpha_{lm}}{\sum_{lms}\alpha_{lm}}
\end{align}
where $\alpha_{lm}$ is the fraction (probability) of ions incorporated in ionic clusters containing $l$ cations and $m$ anions (see SI for details). The phrase ``weight-averaged" arises because $\alpha_{lm}$ is weighted by the total number of ions in the cluster. Of course $\alpha_{lm}$ and thus $\bar{n}_w$ may be computed from MD directly. The theoretical weight-averaged degree of aggregation is derived in the SI, and can be written analytically in closed-form, first derived by Stockmayer in 1952~\cite{stockmayer1952molecular}. Note that in Fig.~\ref{fig:comp}b, we are plotting the weight-averaged degree of average of only the \emph{finite} aggregates in the sol. Thus for the MD measurements, we exclude the largest cluster at each timestep when computing $\alpha_{lm}$. We observe that the simulated weight-average degree of aggregation has a maximum occurring at 19~m, which is in good agreement with the theoretically obtained value of 19.5~m. Moreover, the theoretical and simulated weight-average degrees of aggregation agree very well with each other across the entire concentration range. Though, the MD measurements do not display true divergence at the gel point, like the theoretical prediction, due to the finite size of our simulation~\cite{Christensen2005}. This finite simulation box effectively caps the size that finite ion clusters, preventing the weight-averaged degree of aggregation from truly diverging.

In Fig.~\ref{fig:comp}c we plot the Li$+$ coordination by water (blue curve) and the Li$+$ coordination by anion (red curve) as functions of salt concentration, and we observe a quantitative match of the theory and MD simulations. As we previously mentioned, these curves demonstrate the competition between ion solvation and ion association. The inclusion of this essential physics in our model is responsible for a large majority of its success. It also should be noted that the onset of ionic gelation does not leave any characteristic signature in the coordination numbers. Instead, we observe the coordination numbers of Li$+$ by both water and anion progress smoothly through the critical gel point concentration.

\section{Discussion}

\subsection{Origins of Charged Nano-Domains in WiSEs}

The existence of percolating ionic networks in WiSEs has been noted in literature. For example, in Refs.~\citenum{choi2018,choi2018graph,yu2020asymmetric,andersson2020ion}, the authors performed MD simulations and found percolating ion networks in super-concentrated electrolytes when the ions contained so-called khaotropic anions, such as TFSI$^{-}$. Moreover, in Ref.~\citenum{lewis2020signatures}, vibrational spectroscopy indicated the existence of high order aggregates and inter-connected ion networks in super-concentrated LiTFSI electrolytes. It is clear that our model and simulations are congruent with these observations of percolating ion networks via its prediction of gelation in WiSEs at high enough salt concentrations.

%
However,  in addition to the percolating ion networks, Refs.~\citenum{borodin2017liquid,lim2018,zheng2018understanding,gonzalez2020nanoscale} all discuss the emergence of lithium and water-rich nano-domains, which allow for the facile transport of Li$^+$ within the electrolyte. Interestingly, our model can shed light on the formation of these nano-domains, as well.  Within our theoretical framework, the electrolyte species that are not associated to the gel are referred to as the \emph{sol}, which consists of just free species and finite clusters. We can think about the sol as potentially forming a second network in the electrolyte, if the sol spatially percolates through the electrolyte. In this case, the sol would form a network not via ionic associations, but rather by the virtue of simply occupying enough volume to ensure spatial percolation through the electrolyte. 

To determine if the sol network spatially percolates, a ``site" percolation model can be used, where the electrolyte is divided into lattice-like voxels with some probability of the voxel being occupied by the sol. This probability of a voxel being occupied by the sol is equivalent to the volume fraction of the sol, $\phi_{sol}$ ($\phi_{sol}=1-\phi_{gel}$). It is clear that when $\phi_{sol}$ is large enough (or conversely $\phi_{gel}$ is small enough), the sol will easily percolate through the electrolyte. For example, for salt concentrations below the gel point, the entire electrolyte is sol, and it trivially percolates through the electrolyte. However, when $\phi_{sol}$ falls below a certain threshold, site percolation theory predicts that te sol would no longer percolate through the electrolyte. The critical $\phi_{sol}$ for site percolation is not certain, as our electrolyte does not reside on a crystalline lattice structure. However, critical values for simple lattice structures will range from $\phi^{*}_{sol}=0.2-0.31$ (Simple cubic: 0.31, BCC: 0.25, and FCC: 0.20)~\cite{Christensen2005}. We note that the percolation of sol could also be framed in the context of continuum percolation theory, which would not require the specification of a lattice, but rather the shapes of the percolating species. Still, continuum percolation in three dimensions would yield roughly the same critical percolation thresholds (between 0.19 and 0.3 for spheroids with aspect ratio less than four~\cite{garboczi1995geometrical}). 

Interestingly, this sol ``network" will share many of the same properties of the water and lithium rich nano-domains identified in Refs.~\citenum{borodin2017liquid,lim2018,zheng2018understanding,yu2020asymmetric,gonzalez2020nanoscale}.First, when $f_+>f_-$, the sol will have a net positive charge (as noted in Refs.~\citenum{borodin2017liquid,yu2020asymmetric}) containing more Li$^+$ than anion. Second, the sol will also contain increasingly \emph{free} ionic species as the ionic gel becomes more substantial~\cite{mceldrew2020theory}. Lastly, when these free species are majority Li$^{+}$ (as will be the case for $f_+>f_-$), the model ensures that the free Li$^+$ is fully hydrated by water. These properties, yield a sol network that is water and Li$^+$ rich when the ion gel network is substantial and $f_+>f_-$. However, in order for this water and Li$^+$ rich network to facilitate fast transport of Li$^+$ it must \emph{percolate} through the electrolyte. Otherwise, the sol will exist only in disconnected pockets, and ion transport would likely require more activated processes such as association/dissociation from the gel.

\begin{figure*}[hbt!]
    \centering
    \includegraphics[width=\textwidth]{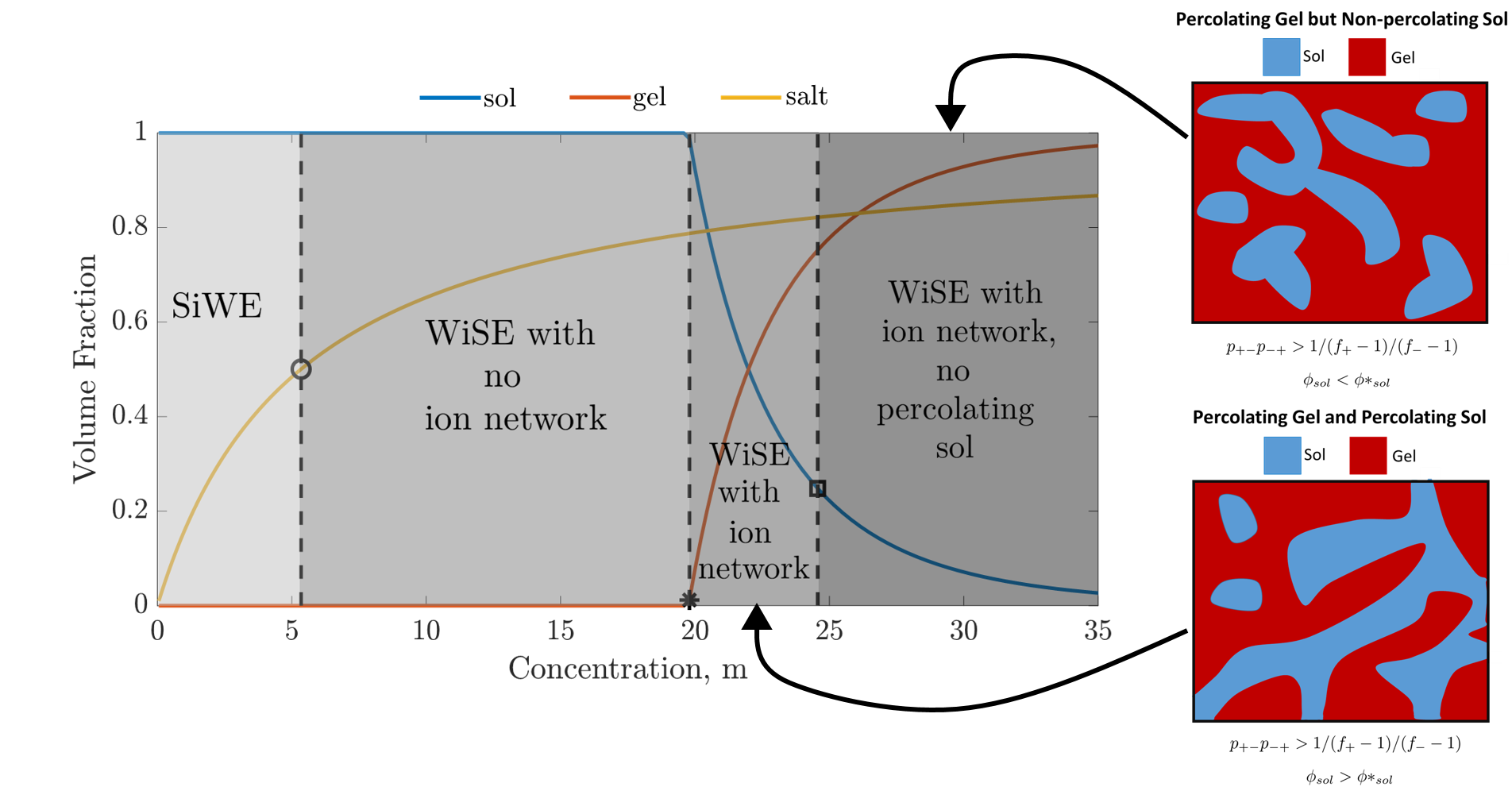}
    \caption{The volume fraction of gel (blue curve), sol (red curve) and total salt (yellow curve) in LiTFSI electrolyte are plotted as a function of salt concentration. Various concentration regimes are depicted. The Salt-in-Water Electrolyte (SiWE) regime spans volume fractions of salt less than 0.5. The Water-in-Salt Electrolyte (WiSE) pre-gel regime spans salt volume fractions greater than 0.5 but concentrations below the critical gel concentration. The WiSE gel regime spans WiSEs with concentrations above the critical gel concentration. WiSEs with a sol volume fraction greater than 0.25 (for BCC lattice) will have a percolating sol phase. WiSEs with a sol volume fraction of less than 0.25 will not have a percolating sol phase. The theoretical curves were generated using the parameters $\lambda=0.075$, $\xi_+=0.4$, $\xi_-=11.3$, $f_+=4$, and $f_-=3$.}
    \label{fig:nano}
\end{figure*}

In Fig.~\ref{fig:nano}, we show several important concentration regimes emerging in LiTFSI. The most dilute region is the Salt-in-Water Electrolyte (SiWE) regime, where the volume fraction of the salt is less than 0.5 (the majority of the volume of the electrolyte is water). For concentrations ranging from 5-20~m, we have a WiSE without a percolating ion gel (the sol will trivially percolate through the electrolyte in this regime). Upon increasing the salt concentration past the critical gel concentration ($\sim$20~m), an ionic gel emerges and grows rapidly in mass and volume. From about 20~m-27.5~m the WiSE exists with both percolating ion networks, as well as percolating sol. For concentrations greater than about 27.5~m, the volume fraction of sol ($1-\phi_{gel}$) reduces below 0.25 (BCC lattice assumed here) and the sol no longer percolates through the electrolyte. In this regime the transport properties of the WiSE could substantially altered, as the pockets of sol will become trapped within the strongly gelled electrolyte. For example, when the sol is able to spatially percolate, we can expect that conduction largely occurs via sol-mediated vehicular transport of ions. When the sol no longer spatially percolates through the electrolyte, ion conduction would likely depend on alternative mechanisms, such as activated hopping mechanisms\cite{andersson2020ion} or ion-exchange between sol and gel phases, similar to the ``ionic semiconductor" concept introduced in ref. \citenum{feng2019free}.

\subsection{Ion transference number at super-concentration}

It is well known that the ionic conductivity of WiSEs is strongly non-monotonic, peaking at intermediate concentrations, before strongly decreasing at higher concentrations~\cite{Suo2015,ding2017conductivity,borodin2017liquid,ding2018phase,horwitz2020mobility}. The reason for this non-monotonicity has been attributed to the increasing viscosity at high salt concentrations, as well as the substantial ionic clustering in WiSEs, which reduces the ``mobile" charge carriers in solution. However, what is less well established is the asymmetry of the ionic transference numbers in WiSEs. It has been observed in LiTFSI electrolytes that the Li$^+$ transference number increases as a function of concentration. This asymmetry has been attributed to the emergence of asymmetric ionic clusters, resulting in an unequal dissociation of Li$^+$ and TFSI$^-$, favoring more free Li$^+$~\cite{borodin2017liquid,yu2020asymmetric,horwitz2020mobility}.

As eluded to in the previous section, our model is able to rationalize the presence of asymmetric clusters in WiSEs, which had not been established clearly in literature. If the cations and anions are able to make a different numbers of associations ($f_+ \neq f_-$), then the associations will be asymmetrically distributed amongst them. Specifically, when $f_+>f_-$, cations will necessarily have a lower ion association probability than anions ($p_{+-}<p_{-+}$), which will inevitable yield a greater number of free cations than anions. This asymmetry is exacerbated when the ionic gel is present, as we described in the previous section. The unequal ion functionalities result in the gel and sol phases with a non-zero, but equal and opposite, net charge. For $f_+>f_-$, this results in a net positive sol, and a net negative gel, as has been observed in literature~\cite{borodin2017liquid,lim2018,zheng2018understanding,yu2020asymmetric,gonzalez2020nanoscale,andersson2020ion}. Moreover, because the ionic gel is percolating and macromolecular, we can assume that it does not have translational degrees of freedom and thus will not contribute to the ionic current. On the other hand, the net positive sol will dominate the ionic current, and yield cationic transference numbers greater than 0.5~\cite{mceldrew2020theory,mceldrew2020corr}. Furthermore, the ion association lifetimes (3-13 ps) indicated that only clusters containing less than 3-5 ions (as estimated by eq. \eqref{eq:constraint}) live long enough to conduct current, with free ions dominating the population of those current-conducting species. When free ions dominate the conductivity~\cite{feng2019free,mceldrew2020theory,mceldrew2020corr}, we can approximate the ionic transference number in terms of the free cation and anion fractions, $\alpha_\pm$, and the cation/anion self-diffusion coefficient ratio, $\mathcal{R}=D_+^0/D_-^0$:
\begin{align}
    t_+ = \frac{\alpha_+ \mathcal{R}}{\alpha_+\mathcal{R} + \alpha_-}.
    \label{eq:tp}
\end{align}
This formula can be used to \textit{estimate} the transference number in WiSEs using computed free ion fractions from theory or from MD simulations, and by knowing the self-diffusion coefficient ratio in the dilute regime, where the free ion fractions of anions and cations will be approximately equivalent.

\begin{figure}[hbt!]
    \centering
    \includegraphics[width=0.5\textwidth]{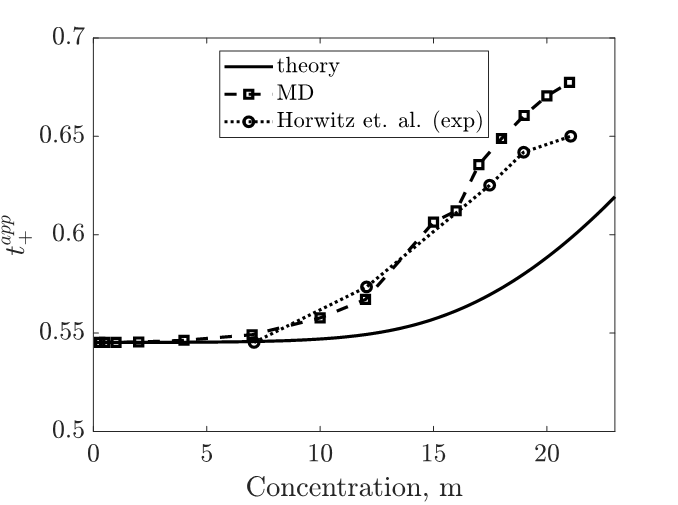}
    \caption{The ``apparent" cationic transference number is plotted as a function of salt concentration for aqueous LiTFSI electrolyte. The theoretical curve (smooth line) and MD data (squares connected by dashed line) were generated from Eq.~\eqref{eq:tp}, using $\mathcal{R}=1.17$, as determined from the experimental data (circles connected by dotted line) from Ref.~\citenum{horwitz2020mobility}. The theoretical curve was generated using the parameters $\lambda=0.075$, $\xi_+=0.4$, $\xi_-=11.3$, $f_+=4$, and $f_-=3$.}
    \label{fig:tpapp}
\end{figure}

Of course the cation transference number may be computed explicitly from the MD simulation via the corrresponding Green-Kubo formula or equivalent mean-square displacement. However, these computations have been performed elsewhere~\cite{borodin2017liquid,gonzalez2020nanoscale}. Thus, for simplicity we use Eq.~\eqref{eq:tp} with free ion fractions  as computed from the theory or MD simulations (note that some more sophisticated equations were derived in Ref.~\citenum{mceldrew2020corr}). The ratio, $\mathcal{R}$ was specified from the experimental data to be $1.17$ by examining the experimental transference number at the lowest salt concentration (where the free cation and anion fractions will be roughly equivalent), as seen in Fig.~\ref{fig:tpapp}. Note, the experimental data is obtained from pulsed field gradient NMR, which does not account for cross-correlations between ions, thus the transference number is denoted as the ``apparent" transference number. Again the MD simulation curve in fig. \ref{fig:tpapp} uses eq. \eqref{eq:tp}, but with free ion fractions computed from MD simulation. Nonetheless, we can see in Fig.~\ref{fig:tpapp} that the MD estimated $t_+$ is very close to the experimental values, which strongly implies that the current conducted predominantly by free species in WiSEs. The theoretical curve qualitatively agrees with the MD and experimental data, but there is appreciable difference between the theoretically predicted transference numbers. This is because the theory actually underpredicts the asymmetry of free cations to free anions. Nonetheless, the qualitative match of the theory suggests uneven ion functionalities are extremely important factors in manipulating the ion transference number.

\subsection{Implications for the Electrochemical Stability Window of WiSEs}

Perhaps the most exciting property of WiSEs, and super-concentrated electrolytes in general, is there expanded electrochemical stability window (ESW). Two primary aspects that govern the ESW of WiSEs is the ability of the WiSE to form a solid-electrolyte interface (SEI) at the negative electrode, and the reduced thermodynamic activity of water in the electrolyte. The formation of the SEI kinetically suppresses the hydrogen evolution at the negative electrode, and the reduced activity of water decreases the thermodynamic driving force for the oxygen evolution on the positive electrode.

The most important aspect of ESWs in WiSEs is its ability to form an SEI layer. In WiSEs, the electrolyte can no longer rely on an organic solvent to mediate and engage in SEI chemistry. Rather, the salt alone must undergo the reduction reactions that produce the SEI~\cite{steinruck2020interfacial}. In this case, the specific chemical make-up of salt (specifically the anion), as well as the precise positioning of species near the electrochemical interface will strongly control the chemistry of SEI formation, as has been discussed at length in literature~\cite{borodin2017modeling,Steinruck2018}. However, one aspect of SEI formation that is scarcely considered is the thermodynamic activity of the salt ions that undergo this chemistry. The activity, $a_i$, of the salt ions will strongly govern the driving force for salt to undergo the SEI-forming chemical reactions. Our model allows us to predict the thermodynamic activity of the salt as a function of the salt concentration, the equations for which are given in the SI. 

\begin{figure}[hbt!]
    \centering
    \includegraphics[width=0.5\textwidth]{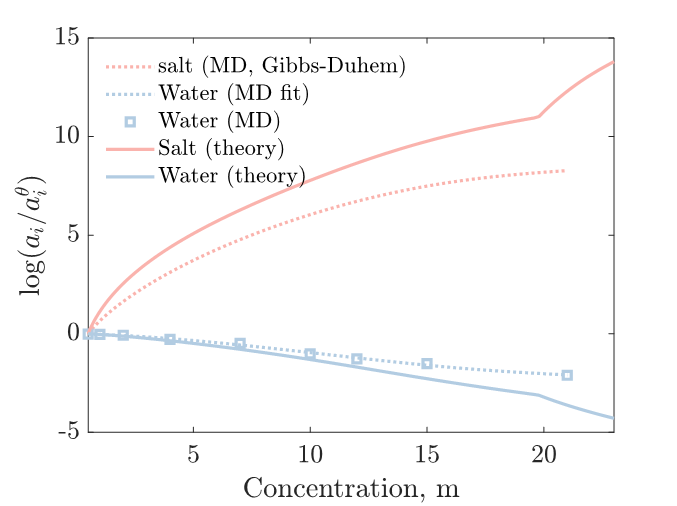}
    \caption{The difference in log of activity of water (blue) and salt (red) from that at 0.2~m are plotted as a function of salt concentration for aqueous LiTFSI electrolyte. Values of water activity are measured from molecular dynamics explicitly (blue squares), and then fitted via a 4~th order polynomial (blue dotted curve). The fitted curve is inserted into the Gibbs-Duhem equation to reconstruct the activity of the salt (red dotted curve). The kinks in the theoretical curves occur at the gel point. The theoretical curves (smooth lines) were generated from Eqs.~(S56), (S58) \& (S59) in the SI, using the parameters $\lambda=0.075$, $\xi_+=0.4$, $\xi_-=11.3$, $f_+=4$, and $f_-=3$. }
    \label{fig:act}
\end{figure}

As we can see in Fig.~\ref{fig:act}, the activity of the LiTFSI salt ($a_{salt}=\sqrt{a_+a_-}$) is predicted by the theory (red smooth line) to increase drastically with increasing salt concentration. In Fig.~\ref{fig:act} we also plot the activity of the LiTFSI salt as calculated from MD simulation. The precise method of for computing the activity of the salt is outlined in the SI. In brief, we explicitly measured the activity of water in the MD simulation (blue squares) via the Osmotic Pressure Activity of Solvent (OPAS) method~\cite{kohns2016solvent}, which is explained in more detail in the SI. Essentially, the activity of the water in an electrolyte of interest can be determined from the osmotic pressure difference between a chamber with pure water and the electrolyte of interest.  We then fit a polynomial curve (blue dotted line) to water activity as a function of salt concentration and integrated the Gibbs-Duhem equation to reconstruct the activity of the salt (red dotted line). We see that our reconstructed salt activity has a strong qualitative agreement with the theoretical prediction. Note, the activities plotted in Fig.~\ref{fig:act} are plotted as a difference from a reference state ($\theta$) of 0.5~m. Of course, this data range does not deeply explore the dilute regime~\cite{debye1923theory,huckel1925theorie}. For very dilute electrolytes ($<$0.5~m), the ionic activity should decrease with concentration. However, for the concentration range studied here, the salt activity increases strongly as a function of salt concentration (at high salt concentrations beyond the Debye-H\"uckel regime) primarily because Li$^+$ starts to become more associated to anions as opposed to solvent. Moreover, the increased clustering reduces the overall translational degrees of freedom of salt in the system. The theory also predicts a kink in chemical potential at the gel point, leading to a more rapid increase in the salt chemical potential, which is not observed in the reconstructed simulation curve. This kink is likely due to the more rapid reduction of translational degrees of freedom of the salt when increasing the salt concentration beyond the critical gel concentration. Note, that we do not observe the this kink in our simulated system. However, this could be due to the fitting process of the water activity, which assumes a smooth transition through the gel point, and thus yields a smooth curve for the Gibbs-Duhem reconstruction  of the salt activity. In order to resolve such a kink in the MD simulations, we would need a much high density of simulation data near the gel point. 

The overall increased salt activity will translate into an increase in the reduction potential of any salt-derived SEI product. In order to demonstrate this fact, consider the following reaction scheme, which was recently determined to be the principle reaction pathway for LiF SEI formation in LiTFSI WiSEs~\cite{steinruck2020interfacial}:
\begin{align}
    \text{Li}^++(\text{R-F})^- + \text{e}^- \rightleftharpoons \text{LiF}+ R^-
\end{align}
where for a LiTFSI electrolyte R=CF$_3$-SO$_2$-N-SO$_2$-CF$_2$. The equilibrium reduction potential, $\Delta \Phi^{eq}$, is given by the Nernst equation
\begin{align}
    \Delta \Phi^{eq}=E^{\theta}+k_bT\ln\left(\frac{a_{\text{Li}^{+}}a_{(\text{R-F})^-}}{a_{\text{LiF}}a_{\text{R}^-}}\right)
\end{align}
where 
\begin{align}
    E^{\theta}=\frac{\mu^{\theta}_{\text{Li}^{+}}\mu^{\theta}_{(\text{R-F})^-}}{\mu^{\theta}_{\text{LiF}}\mu^{\theta}_{\text{R}^-}}
\end{align} 
and $\mu^{\theta}_i$ is a reference chemical potential of species $i$ (assumed to be the standard dilute chemical potential for electrolytic species). Because LiF is deposited on the electrode surface as a separate phase, we can assume that $a_{\text{LiF}}$ is not \emph{greatly} affected by electrolyte salt concentration and that $a_{\text{LiF}} \approx 1$. Our model cannot definitively say much about the activity R$^-$, but for simplicity we approximate $a_{\text{R}^-} \approx 1$. In this case, the equilibrium reduction potential for the LiF SEI formation will be shifted by the following factor:
\begin{align}
    \Delta \Phi^{eq}-E^{\theta}=k_bT \ln \left(a_{\text{Li}^{+}}a_{(\text{R-F})^-}\right)= 2 k_bT \ln a_{salt}
\end{align}
Thus, we can estimate that the increase in salt activity (from 0.2~m to 21~m) would theoretically result in a positive shift of 0.69~eV (0.44~eV for MD) toward higher electrode potentials, resulting in a strongly increased affinity to form the SEI layer. In fact, this increase in affinity may explain why purely-salt derived SEI layers require extreme salt concentrations: it is simply thermodynamically unfavorable for the salt to be reduced at low salt concentrations. Thus, understanding how the chemical potential of the salt varies as a function of salt concentration may be a key consideration when designing new WiSEs. 

It has also been well-established that the coordination environment of Li$^+$ strongly affects the SEI chemistry~\cite{xu2007solvation,xu2007charge,von2012correlating}. For example, in Ref.~\citenum{von2012correlating}, von Wald Cresce \textit{et al.} studied an organic ethylene carbonate (EC)/propylene carbonate (PC) electrolyte linking the preference of PC over EC to be involved in SEI formation to the preference of PC over EC to be in Li$^+$ solvation sheath. In 21~m LiTFSI, the average coordination environment of Li$^+$ contains roughly two anions and two water molecules, as opposed to the typical pristine four-water solvation shell in dilute solutions. Thus, the direct contact between Li$^+$ and the TFSI$^-$ is thought to enable their combined reduction to the LiF-rich SEI layer~\cite{Suo2015}. This is consistent with the thermodynamic explanation based on salt activity presented above. 

Furthermore, for a simple 3 component (cation, anion and solvent) electrolyte the SEI chemistry that is possible is narrow, and can perhaps be grasped intuitively without too much theoretical guidance. However, the next generation of battery electrolytes may contain multiple solvents \cite{wang2018hybrid,zhang2018aqueous,dou2018safe,Dou2019,molinari2020chelation} or multiple salts~\cite{lui2011salts,Suo2016,kondou2018enhanced,molinari2019transport,molinari2019general,chen202063,jiang2020high,becker2020hybrid}. In this case, predictive, chemical-specific models will be extremely important in building our intuition. Knowing how the Li$^+$ coordination environment changes for different complex, super-concentrated salt/solvent blends will be extremely non-trivial, and will help aid in understanding increasingly complex SEI chemistry. Our current model, is equipped to handle the simple 3 component WiSEs, but further development will be needed to generalize it to these more complex mixtures. 

Another aspect that can affect the electrochemical stability of WiSEs is the activity of water, which factors strongly into the ability for the water to be stabilized in the presence of large voltages. The activity of water is expected to decrease with increasing salt concentration, as predicted in Fig.~\ref{fig:act} by both our model (smooth line) and the simulation (smooth line) for LiTFSI electrolyte. This is a result of water transitioning from being primarily unbound to Li$^{+}$ at low salt concentrations, to being primarily bound to Li$^+$ at high salt concentrations. The Li-water interaction is strongly energetically favorable and reduces the activity of water. The Nernst equation again implies that activity of water will tend to shift the oxidation potential of water by a factor of $-4k_BT \log(a_w)$, corresponding to a theoretical 0.37~eV (0.21~eV from MD) shift in potential from a concentration of 0.2~m to a concentration of 21~m. This implies that when the activity of water is reduced below 1, the equilibrium potential for the oxygen evolution reaction is shifted to more positive potentials. Additionally, it was also observed that water molecules are considerably depleted from positive electrode surfaces \cite{vatamanu2017,mceldrew2018} in super-concentrated electrolytes. Thus, oxygen evolution reaction would be expected to be suppressed even further than what was thermodynamically predicted here.

Of course electrochemical window may not be considered only in terms of chemical thermodynamics. There are kinetic factors that will determine the reactivity of water or the formation of SEI at active electrodes, such as e.g. electron transfer reorganization energy, local populations of reactive species, or electrocatalytic effects involving adsorbed intermediates. Obviously such considerations lie beyond the scope of the presented  analysis, but even the 'classical' concepts of activities discussed above, give a qualitative indication of the trend in water reactivity and SEI formation with the increase of the salt concentration, in line with available experimental observations.

\section{Conclusion}

In this article we have outlined the theory of ion aggregation and gelation in water-in-salt electrolytes. Our model contains a small number of physically transparent molecular parameters, which can be readily obtained from molecular simulations. The model was validated across an extremely large concentration range from dilute (0.2~m) to super-concentrated (21~m) for aqueous LiTFSI, LiOTF, and LiFSI electrolytes. In our molecular simulations we observed extensive ion clustering and the emergence of a percolating ion network, which our model was able to represent as an ionic gel. The model is able to account for the strong competition between ionic association and ion solvation, as was observed in the performed simulations. In so doing, we are able to reproduce the ion cluster distributions, Li$^+$ coordination, and the degree of ion aggregation as a function of salt concentration for all of the simulated electrolytes. 

Moreover, our model seems to be able to help explain various intriguing properties of WiSE's that have been observed in the literature. In particular, there have been various observations detailing the heterogeneous nano-structure in WiSEs, specifically the emergence of percolating ion networks interpenetrated by water/Li$^+$. Our theory is able to predict the conditions required for such structures to emerge, as well as the general trend of asymmetric clusters leading to cationic transference numbers increasing considerably above 0.5 as salt concentration increases. Additionally, we were able to use a thermodynamic argument, involving the simultaneous increase of salt activity and decrease water activity at high salt concentrations, to explain the preference of SEI formation over hydrogen evolution. This preference of SEI formation is essential in extending the electrochemical stability window of water-in-salt electrolytes. Thus, understanding this process, may unlock the design principles for further optimization of WiSEs.

\section{Acknowledgements}

All authors would like to acknowledge the Imperial College-MIT seed fund. M.M. and M.Z.B. acknowledge support from a Amar G. Bose Research Grant. Z.A.H.G was supported through a studentship in the Centre for Doctoral Training on Theory and Simulation of Materials at Imperial College London funded by the EPSRC (EP/L015579/1) and from the Thomas Young Centre under grant number TYC-101. A.A.K would like to acknowledge the research grant by the Leverhulme Trust (RPG-2016- 223). This work used the Extreme Science and Engineering Discovery Environment (XSEDE), which is supported by National Science Foundation grant number ACI-1548562. S.B. was supported by the National Natural Science Foundation of China (51876072) and the financial support from the China Scholarship Council.

\section{Appendix}

\subsection{Mixture Stoichiometry}
As described in the main chapter, we consider a polydisperse mixture of $\sum_{lms}N_{lms}$ ionic clusters, each containing $l$ cations, $m$ anions, and $s$ water molecules associated to cations (rank $lms$ cluster), and (if present) an interpenetrating gel network containing $N_+^{gel}$ cations, $N_-^{gel}$ anions, and $N_0^{gel}$ water molecules. We model the cations to have a functionality (defined as the number of associations that the species can make) of $f_+$, and anions to have a functionality of $f_-$. This means that a cation is able to associate with $f_+$  anions or water molecules. Likewise, anions may associate  with up to $f_-$ cations. Water molecules may associate (hydrate) to a single cation (functionality of one).
 
Our model treats the electrolyte as lattice fluid. We designate a single lattice site to have the volume of a single water molecule, $v_0$.  Thus, the entire volume of the mixture, $V$, is divided into $\Omega = V/v_0$ lattice sites. Moreover, cations will occupy $\xi_+=v_+/v_0$ lattice sites, and anions will occupy $\xi_-=v_-/v_0$ lattice sites. Furthermore, when a gel is formed, we distinguish between the volume fractions of gel (superscript $gel$) and sol (superscript $sol$). The volume fractions in the sol and gel constitutes the total volume fraction, $\phi_j$ of a given species, $j$, is given by
\begin{align}
    \phi_j=\phi_j^{sol}+\phi_j^{gel}    
\end{align}
in which the gel volume fractions is defined as $\phi_j^{gel}=\xi_j N_j^{gel}/\Omega$, with $N_j^{gel}$ as the mole number of species $j$ in the gel. The subscript $j=+,-,0$ corresponds to cation, anion, and solvent, respectively. The sol volume fraction of cations, anions, and solvent molecules have, respectively, the definitions
\begin{align}
    \phi^{sol}_+=\sum_{lms} \xi_+l c_{lms}
\end{align}
\begin{align}
    \phi^{sol}_-=\sum_{lms} \xi_-
    m c_{lms}
\end{align}
\begin{align}
    \phi^{sol}_0=\sum_{lms} s c_{lms}
\end{align}
where $c_{lms}=N_{lms}/\Omega$ is the dimensionless concentration of a $lms$ cluster (the number of $lms$ clusters per lattice site). Similarly, we define $\phi_\pm=\phi_++\phi_-$, which is the total volume fraction of the salt in solution. For simplicity the mixture is assumed to be incompressible, i.e.
\begin{align}
    1=\phi_\pm+\phi_0=\phi_++\phi_-+\phi_0
    \label{eq:incomp}
\end{align}
$\phi_+$ and $\phi_-$ are not independent owing to electroneutrality: $\phi_+/\xi_+=\phi_-/\xi_-$. The reduced volume of the mixture, $\Omega$, can also be expressed in terms of the mole number of each species/component due to the incompressibility constraint [Eq.~\eqref{eq:incomp}]
\begin{align}
    \Omega=\sum_{lms}(\xi_+l+\xi_-m+s)N_{lmsq}+\xi_+N^{gel}_++\xi_-N^{gel}_-+N^{gel}_0
    \label{eq:O}
\end{align}
This definition must be used when differentiating the free energy of mixture. 

\subsection{Pre-gel Cluster Distribution and Association Probabilities}
Recall the the free energy of mixing, $\Delta F$, written in the main text:
\begin{align}
    \beta \Delta F &= \sum_{l,m,s} \left[N_{lms}\log \left( \phi_{lms} \right)+N_{lms}\Delta_{lms}\right] \nonumber \\
    &+ \Delta^{gel}_+ N^{gel}_+ + \Delta^{gel}_- N^{gel}_- + \Delta^{gel}_0 N^{gel}_0 
    \label{eq:FSI}
\end{align}
where $\beta=1/k_BT$ is the inverse thermal energy, and $\Delta_{lms}$ is the free energy of formation for clusters of rank $lms$, and $\Delta^{gel}_i$ is the free energy change of species $i$ upon association to the gel.

We may differentiate the free energy with respect to $N_{lms}$ to obtain the chemical potential of a cluster rank $lms$
\begin{align}
    \beta \mu_{lms}&=\ln \phi_{lms} + 1 - (\xi_+ l + \xi_-m+s) c_{tot}+\Delta_{lms} \nonumber \\
    &+d\left[\phi_0 (\xi_+l+\xi_-m+s)-s\right],
     \label{eq:muclust}
\end{align}
where $c_{tot}=\sum_{lms}c_{lms}$ is the total reduced concentration, and $d=\Delta^{gel\prime}_+c_+^{gel}+\Delta^{gel\prime}_-c_-^{gel}+\Delta^{gel\prime}_0c_0^{gel}$ (the $\prime$ notation refers to a derivative with respect to $\phi_\pm$), and  $c_i^{gel}=N_i^{gel}/\Omega$ is dimensionless concentration of $i$ in the gel (number of species $i$ in the gel per lattice site). Additionally, we may define the chemical potential of species immersed in the gel, which yields
\begin{align}
    \beta \mu_+^{gel}=\Delta^{gel}_+  - \xi_+c_{tot} +\xi_+d\phi_0
\end{align}
for cations,
\begin{align}
    \beta \mu_-^{gel}=\Delta^{gel}_--\xi_-c_{tot}+\xi_-d\phi_0
\end{align}
for anions, and 
\begin{align}
    \beta \mu_0^{gel}=\Delta^{gel}_0-c_{tot}+d\phi_\pm
\end{align}
for water molecules.

The distribution of clusters can be derived by enforcing a chemical equilibrium between all of the clusters and their bare constituents (unassociated components)
\begin{align}
    l [\text{bare cation}]+m [\text{bare anion}]+(s+q) [\text{bare solvent}]\rightleftharpoons [lmsq\,\,\text{cluster}]
\end{align}
Chemical equilibrium requires that the chemical potentials of bare species and those in clusters are equivalent
\begin{align}
    l \mu_{100}+m \mu_{010}+s\mu_{001} = \mu_{lms} = l\mu^+_{lms} + m\mu^-_{lms} + s\mu^{0}_{lms}.
    \label{eq:eqm}
\end{align}
In Eq.~\eqref{eq:eqm}, we have defined the chemical potential of a cation, anion or solvent molecule in an arbitrary cluster in the following manner
\begin{align}
    \mu^+_{lms}=\frac{\partial \mu_{lms}}{\partial l}=\mu_{100}
\end{align}
\begin{align}
    \mu^-_{lms}=\frac{\partial \mu_{lms}}{\partial m}=\mu_{010}
\end{align}
and
\begin{align}
    \mu^{0}_{lms}=\frac{\partial \mu_{lms}}{\partial s}=\mu_{001}
\end{align}
Thus, it becomes clear that the equilibrium condition in Eq.~\eqref{eq:eqm} requires that species within an arbitrary cluster ($\mu^{i}_{lms}$, $i=+,-,0$) have equivalent chemical potentials to the bare species not in the cluster ($\mu_{100}$, $\mu_{010}$, $\mu_{001}$).

We may solve Eq.~\eqref{eq:eqm} for an arbitrary cluster rank $lms$ obtaining the following relationship:
\begin{align}
    \phi_{lms}=K_{lms}\phi_{100}^l \phi_{010}^m \phi_{001}^s
    \label{eq:cluster_distSI}
\end{align}
where $K_{lms}$ is the equilibrium constant with the following definition:
\begin{align}
    K_{lms}=\exp(l+m+s-1-\Delta_{lms})
\end{align}
As we mentioned in the main text, the number of water molecules in the cluster ($s$) is specified by the number of cations ($l$) and the number of anions ($m$) in the cluster: $s=f_+l-l-m+1$. Thus, strictly speaking, the subscripts $lms$ describing the rank of a cluster for any variable (such as $\Delta_{lms}$) could simply be described by the subscript $lm$. However, for book-keeping purposed we will keep the subscript $lms$ throughout. The partitioning of the species into clusters of different sizes is determined almost entirely from $\Delta_{lms}$. As such, this is where much of the physics of the ion/solvent association will be included. $\Delta_{lms}$ contains three major contributions: 1) \emph{combinatorial} describing the multiplicity of clusters with the same number of constituents, 2) \emph{binding} describing the association enthalpy of the constituents in the cluster, and 3) \emph{configurational} describing the configurational entropy change upon forming a cluster from base constituents. As previously mentioned, in the sticky-cation approximation we have neglected the non-ideal electrostatic contributions to $\Delta_{lms}$, such as Debye-H\"uckle like screening or Born solvation free energy. Thus,
\begin{align}
    \Delta_{lms}=\Delta_{lms}^{comb}+\Delta_{lms}^{bind}+\Delta_{lms}^{conf}.
\end{align}

The combinatorial contribution $\Delta^{comb}_{lms}$ is a purely entropic contribution. Thus, it requires the combinatorial enumeration, $W_{lms}$, of all of the possible ways a cluster with $l$ cations, $m$ anions, and $s$ solvent molecules can be formed:
\begin{align}
    \Delta_{lms}^{comb}=-\log \left(W_{lms}\right)
    \label{eq:dcomb}
\end{align}
Here $W_{lms}$ can be derived in a two step procedure. First, we enumerate the number of ways, $W_{lm}$ to construct a network containing $l$ anions and $m$ cations, which are associated together in an alternating fashion. This combinatorial problem is well known\cite{stockmayer1952molecular}:
\begin{align}
    W_{lm}=\frac{(f_+ l -l)!(f_- m -m)!}{l!m!(f_+l-l-m+1)!(f_-m-l-m+1)!}.
\end{align}
In the second step, we enumerate the number of ways $s$ solvent molecules can be placed on the cation-anion cluster. We know that we may only place the $s$ solvent molecules on the remaining $f_+l-l-m+1$ open cation sites, and actually because there will be no open sites on the cations, and the solvent molecules are indistinguishable, the cluster  will contain $s=f_+l-l-m+1$ solvent molecules, with only one unique  arrangement. Thus, $W_{lms}=W_{lm}$. 

Next, the binding contribution, $\Delta^{bind}_{lms}$, can be describe simply via the association free energies: $\Delta u_{+0}$ for cation-solvent association and $\Delta u_{+-}$ for cation-anion association. For $l+m>0$, the association free energy for an $lms$ cluster is
\begin{align}
    \Delta^{bind}_{lms}=s\Delta u_{+0}+(l+m-1)\Delta u_{+-}
    \label{eq:dbond}
\end{align}

The coefficient in front of $\Delta u_{+-}$ is due to the fact that there must be that many cation/anion associations to form a cluster with $l$ cations and $m$ anions. 
Again $s=f_+l-l-m+1$, thus 
\begin{align}
    \Delta^{bind}_{lms}=f_+l\Delta u_{+0}+(l+m-1)(\Delta u_{+-}-\Delta u_{+0})
\end{align}
For $l+m=0$, $s=1$ (free solvent molecule), we the binding energy is zero:
\begin{align}
    \Delta^{bind}_{lms}=0.    
\end{align}
Recall, that our model does not allow for solvent molecules to form clusters among themselves. For this reason, if a cluster contains 0 cations and anions, the cluster will necessarily only contain a single solvent molecule, corresponding to a free solvent molecule. Clearly, a free water molecule does not form associations and thus $\Delta^{bind}_{001}=0$. Overall, we can write $\Delta^{bind}_{lms}$ as
\begin{align}
    \Delta^{bind}_{lms}=\left[f_+l\Delta u_{+0}+(l+m-1)(\Delta u_{+-}-\Delta u_{+0})\right](1-\delta_{l,0}\delta_{m,0})
\end{align}
where $\delta_{i,j}$ is the Kroenecker delta function. 

Finally, for the configurational contribution, $\Delta^{conf}_{lms}$, as was noted in ref. \citenum{mceldrew2020corr}, we may model the associations as being semi-flexible. In order to do so, associations are partitioned between trans and gauche orientations, which  differ by an energy (units $k_BT$) of $\epsilon$ (trans conformation has the reference energy of 0, and gauche $\epsilon$).
Flory's expression for the entropy of disorientation  used in lattice fluid  theory\cite{flory1942thermodynamics,flory1953principles,tanaka1989,tanaka1999} may be augmented to account for the partitioning of associations into trans and gauche orientations \cite{flory1956statistical,tanaka2011polymer}:
\begin{align}
    \Delta_{lm}^{conf} =-\ln\left(\frac{(\xi_+l+\xi_-m+s)\left[\frac{(Z-2)^{2g}}{Z2eg^{2g}(1-g)^{2(1-g)}}\right]^{l+m+s-1}}{\xi_+^l\xi_-^m}\right)+g\epsilon
\end{align}
where $Z$ is the coordination number of the lattice, and $g$ is the fraction of associations in the gauche conformation given by\cite{flory1956statistical}
\begin{align}
    g=\frac{(Z-2)e^{-\epsilon}}{1+(Z-2)e^{-\epsilon}}
\end{align}
In the high temperature limit ($\epsilon \rightarrow 0$), $g\rightarrow (Z-2)/(Z-1)$ for completely flexible chains. This yield's Flory's original entropy of disorientation result\cite{flory1942thermodynamics}. In the low temperature limit ($\epsilon\rightarrow\infty$), we obtain $g \rightarrow 0$, and all associations are stuck in the trans orientation, thus associations become completely inflexible. As we will see, when working within the Sticky-cation approximation, the configurational entropy actually gets cancelled out, and whether the associations are fully flexible, semi-flexible, or completely inflexible does not matter. 

Plugging in each contribution of $\Delta_{lms}$ into Eq.~\eqref{eq:cluster_distSI}, we obtain the distribution (for $l+m>1$)
\begin{align}
     \tilde{c}_{lms}=\frac{W_{lm}}{\lambda_{+-}} 
    \left(\frac{f_+\phi_{100}\lambda_{+-}}{\xi_+}\right)^l \left(\frac{f_-\phi_{010}\lambda_{+-}}{\xi_-}\right)^m \left(\phi_{001}\lambda_{+0}\right)^{f_+l-l-m+1}
    \label{eq:dist}
\end{align}
where $\lambda_{+-}$ and $\lambda_{+0}$ are the association constants for cation-anion association and cation-solvent association, respectively. They are defined generally as
\begin{align}
    \lambda_{+-}=\frac{(Z-2)^{2g}}{Z g^{2g} (1-g)^{2(1-g)}}\exp \left( -\Delta u_{ij}-g\epsilon \right)
    \label{eq:Lsflex}
\end{align}
where $i=+$  and $j=-,0$ are indices denoting either cation-anion or cation-solvent associations. For completely flexible associations, $g\rightarrow (Z-2)/(Z-1)$ yielding 
\begin{align}
    \lambda^{flex}_{ij}=\frac{(Z-1)^{2}}{Z}\exp \left( -\Delta u_{ij}\right).
    \label{eq:Lflex}
\end{align}
For completely inflexible associations, $g\rightarrow 0$ yielding 
\begin{align}
    \lambda^{inflex}_{ij}=\frac{1}{Z}\exp \left( -\Delta u_{ij}\right).
    \label{eq:Linflex}
\end{align}

Note that in Eq.~\eqref{eq:dist}, we have removed the $s$, index, as the cluster is now defined simply by the number of anions and cations  (recall $s=f_+l-l-m+1$). Furthermore, Eq.~\eqref{eq:dist} gives the thermodynamically consistent number distribution of clusters in terms of the volume fraction of the \emph{bare} cations, anions, and solvent molecules are known, which are generally unknown. 

We may determine the bare species volume fractions by introducing the association probabilities, $p_{ij}$, that an association site of species $i$ is occupied with an association to species $j$. These probabilities are useful because we may write the reduced bare species' volume fractions in terms of these probabilities. For the bare cation volume fraction we have,
\begin{equation}
    \phi_{100}= \phi_+(1-p_{+-}-p_{+0})^{f_+}.
    \label{eq:x}
\end{equation}
The above equation arises because the probability that a given cation association site will be `dangling' (not participating in associations) will be $1-p_{+-}-p_{+0}$. Thus for all $f_+$ sites to be dangling is $(1-p_{+-}-p_{+0})^{f_+}$. Similarly, for the bare anion volume fraction we have
\begin{equation}
    \phi_{010}=\phi_-(1-p_{-+})^{f_-},
    \label{eq:y}
\end{equation}
where $(1-p_{-+})^{f_-}$ is the probability that all $f_-$ anion sites will be unassociated. Finally, for the bare solvent volume fraction we have
\begin{equation}
    \phi_{001}=\phi_0(1-p_{0+}),
    \label{eq:z}
\end{equation}
where $1-p_{0+}$ is the probability that a solvent molecule's single association site will not participate in an association. The probabilities are related in the following manner due to conservation of associations:
\begin{align}
    f_+ \phi_+ p_{+-}/\xi_+=f_- \phi_- p_{-+}/\xi_-=\zeta
    \label{eq:sys1}
\end{align}
where $\zeta$ is the total number of anion-cation associations. 
\begin{align}
    f_+ \phi_+ p_{+0} /\xi_+=\phi_0 p_{0+}=\Gamma
    \label{eq:sys2}
\end{align}
where $\Gamma$ is the total number of cation-solvent associations. 
Tanaka also employs the laws of mass action on the number of associations using the association constants $\lambda_{+-}$ and $\lambda_{+0}$:
\begin{align}
    \lambda_{+-}\zeta=\frac{p_{+-}p_{-+}}{(1-p_{+-}-p_{+0})(1-p_{-+})}.
    \label{eq:sys3}
\end{align}
Similarly, for the cation-solvent associations we have
\begin{align}
    \lambda_{+0}\Gamma=\frac{p_{+0}p_{0+}}{(1-p_{+-}-p_{+0})(1-p_{0+})}
    \label{eq:sys4}    
\end{align}
Here we are essentially treating $\lambda_{+-}$ and $\lambda_{+0}$ as equilibrium constants for individual associations made. However, Eqs.~\eqref{eq:sys3} \& \eqref{eq:sys4} are actually singular, because the ``stickyness" of the cation dictates that there will be no dangling sites on the cation:
\begin{align}
    1=p_{+-}+p_{+0}
    \label{eq:sys3real}
\end{align}
Therefore, we must divide Eqs.~\eqref{eq:sys3} \& \eqref{eq:sys4} to remove the singularity:
\begin{align}
    \tilde{\lambda}=\frac{\lambda_{+-}}{\lambda_{+0}}=\frac{p_{-+}(1-p_{0+})}{p_{0+}(1-p_{-+})}.
    \label{eq:sys4real}
\end{align}
where $\tilde{\lambda}=\exp(-\Delta u_{+-}+\Delta) u_{+0}$, which is independent of flexibility of associations discussed above. 

The Eqs.~\eqref{eq:sys1}-\eqref{eq:sys2} \& Eqs.~\eqref{eq:sys3real}-\eqref{eq:sys4real} provide a system from which we may solve for $p_{+-}$, $p_{-+}$, $p_{0+}$, and $p_{+0}$ in terms of the overall species volume fractions. The solutions are 
\begin{align}
    p_{+-}&=\frac{\psi_0-\phi_++\Tilde{\lambda}(\psi_++\psi_-)}{2(\Tilde{\lambda}-1)\psi_+} \nonumber \\
    &-\frac{\sqrt{4(\Tilde{\lambda}-1)+(\Tilde{\lambda}(\psi_--\psi_+)+\psi_++\psi_0)^2}}{2(\Tilde{\lambda}-1)\psi_+},
    \label{eq:p+-}
\end{align}
\begin{align}
    p_{-+}&=\frac{\psi_0-\phi_++\Tilde{\lambda}(\psi_++\psi_-)}{2(\Tilde{\lambda}-1)\psi_-} \nonumber \\
    &-\frac{\sqrt{4(\Tilde{\lambda}-1)+(\Tilde{\lambda}(\psi_--\psi_+)+\psi_++\psi_0)^2}}{2(\Tilde{\lambda}-1)\psi_-},
\end{align}
\begin{align}
    p_{+0}&=1-\frac{\psi_0-\phi_++\Tilde{\lambda}(\psi_++\psi_-)}{2(\Tilde{\lambda}-1)\psi_+} \nonumber \\
    &+\frac{\sqrt{4(\Tilde{\lambda}-1)+(\Tilde{\lambda}(\psi_--\psi_+)+\psi_++\psi_0)^2}}{2(\Tilde{\lambda}-1)\psi_+},
\end{align}
and 
\begin{align}
    p_{0+}&=\frac{\psi_0}{\psi_+}-\frac{\psi_0-\phi_++\Tilde{\lambda}(\psi_++\psi_-)}{2(\Tilde{\lambda}-1)\psi_0} \nonumber \\
    &+\frac{\sqrt{4(\Tilde{\lambda}-1)+(\Tilde{\lambda}(\psi_--\psi_+)+\psi_++\psi_0)^2}}{2(\Tilde{\lambda}-1)\psi_0},
    \label{eq:p0+}
\end{align}
where $\psi_i=f_i \phi_i/\xi_i=f_i\tilde{c}_i$. Note that $\xi_0=f_0=1$. 
 
Finally, we should rewrite the cluster distribution equation [Eq.~\eqref{eq:dist}], because the current cluster concentration distribution is written in terms of the individual association constants ($\lambda_{+-}$ and $\lambda_{+0}$), which will be singular and the ``bare" cation volume fraction ($\phi_{100}$) which will be zero. We may use Eqs.~\eqref{eq:x}-\eqref{eq:sys4real}, to remove the singularities from the $c_{lm}$ in Eq.~\eqref{eq:dist}. We obtain the following result:
\begin{align}
    \tilde{c}_{lms}=\frac{\psi_0\alpha_0 W_{lm}}{\tilde{\lambda}}\left(\tilde{\lambda} \frac{\psi_+\alpha_+}{\psi_0\alpha_0}\right)^{l}\left(\tilde{\lambda} \frac{\psi_-\alpha_-}{\psi_0\alpha_0}\right)^{m}(\tilde{\lambda}-\delta_{l,0}\delta_{m,0})
\end{align}
where $\alpha_0=1-p_{0+}$, $\alpha_+=(1-p_{+-})^{f_+}$ and $\alpha_-=(1-p_{-+})^{f_-}$ are the fraction of free water molecules, cations and anions, respectively. Note that $\tilde{c}_{00}$ corresponds to the dimensionless concentration of free water molecules. Equivalently, we may write the cluster concentration distribution in terms of the ionic association probabilities:
\begin{align}
    \tilde{c}_{lm}=\tilde{c}_{salt}W_{lm}\mathcal{K}\left(\frac{p_{-+}}{(1-p_{-+})}(1-p_{+-})^{f_+-1}\right)^{l}\left(\frac{p_{+-}}{(1-p_{+-})}(1-p_{-+})^{f_--1}\right)^{m}(\tilde{\lambda}-\delta_{l,0}\delta_{m,0})
\end{align}
where $\tilde{c}_{salt}=\phi_+/\xi_+=\phi_-/\xi_-$ is the dimensionless concentration of salt (\# of salt molecules per lattice site), and $\mathcal{K}=f_\pm(1-p_{-+})(1-p_{+-})/p_{\mp\pm}$. We may divide the distribution by $2\tilde{c}_{salt}$ to obtain the cluster \emph{probability} distribution, $\alpha_{lm}$:
\begin{align}
    \alpha_{lm}=W_{lm}\frac{\mathcal{K}}{2}\left(\frac{p_{-+}}{(1-p_{-+})}(1-p_{+-})^{f_+}\right)^{l}\left(\frac{p_{+-}}{(1-p_{+-})}(1-p_{-+})^{f_-}\right)^{m}(\tilde{\lambda}-\delta_{l,0}\delta_{m,0})
\end{align}
which was similarly defined in Ref.~\citenum{mceldrew2020theory}. 

We also note that the reduced association constant may now be written in the following manner:
\begin{align}
    \tilde{\lambda}=\exp\left(\Delta u_{+-}-\Delta u_{+0}\right)=\exp\left(\beta(\Delta U_{+-}-\Delta U_{+0})\right)
\end{align}
Thus, we can see that the entropic portion of the association constant actually cancels out here, and is determined strictly by the  energetics of association. This, relies on the assumption that the entropy of solvent-cation association is equivalent to cation-anion association. This assumption is somewhat rough, and results from the Flory lattice configurational entropy. In order to relax this assumption, we would have to modify Flory's formulae. For our purposes, however, this is unnecessary, as we see a satisfactory agreement of the model with the simulations. 

\subsection{More on the ``Sticky" Cation Approximation}
\begin{figure*}[hbt!]
    \centering
    \includegraphics[width=1\textwidth]{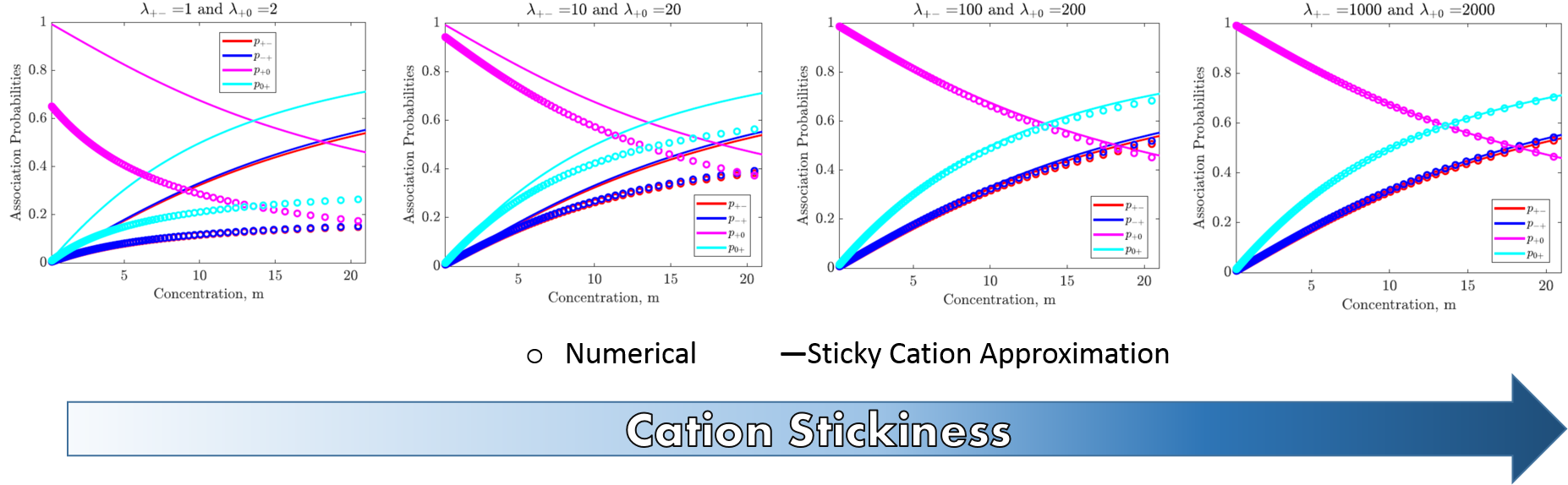}
    \caption{A comparison of the association probabilities using the full numerical solution to eqs. \eqref{eq:sys1}-\eqref{eq:sys4} and the analytical association probabilities (eqs. \eqref{eq:p+-}-\eqref{eq:p0+}) obtained via the ``sticky" cation approximation (SCA). In all panels, the association probabilities using the SCA correspond to $\tilde{\lambda}=1/2$. The association constants, $\lambda_{+-}$ and $\lambda_{+0}$, used to generate the numerical association probabilities are given in the figures, and have an increasing magnitude going from the left to right panels with the same relative ratio.}
    \label{fig:sticky}
\end{figure*}

In order to probe the validity of the sticky cation approximation (SCA), we may compare them to the general association probabilities derived from Eqs.~\eqref{eq:sys1}-\eqref{eq:sys4}, which may be solved numerically. In Fig.~\ref{fig:sticky}, we show those comparisons for different magnitudes of $\lambda_{+-}$ and $\lambda_{+0}$, but with the same relative magnitude, thus keeping a constant $\tilde{\lambda}$ of 1/2. We see that when association constants are of order 10$^2$, the SCA matches the numerical solutions almost exactly. Thus, we see clearly that this approximation is valid for large association constants.

\subsection{Critical Gel Line}
In the main text, we derived the expression for the critical gel line (the critical gel concentration as a function of $\tilde{\lambda}$ or vice versa). In principle, the association constant, $\tilde{\lambda}$, can be mapped to the temperature of the electrolyte. However, it is important to note that when we take a detailed  look at the temperature dependence, we must also question the applicability of the ``Sticky-Cation Approximation" (SCA) across the studied temperature range. As we mentioned in the previous section, in order for the SCA to be appropriate, both $\lambda_{+-}$ and $\lambda_{+0}$ need to be roughly 100 or more.  However, we know that in general, association constants will decrease exponentially as a function of temperature according to the following relation:
\begin{align}
    \lambda_{ij}=\exp\left(- \frac{\Delta F_{ij}}{k_bT}\right)=\exp\left(- \frac{\Delta U_{ij}-T\Delta S_{ij}}{k_bT}\right),
    \label{eq:LL}
\end{align}
where $\Delta F_{ij}$, $\Delta U_{ij}$, and $\Delta S_{ij}$ are respectively the free energy, internal energy, and entropy of association between species $i$ and $j$. Thus, it is clear that raising the temperature, could bring us outside of conditions in which the SCA is appropriate. Furthermore, the partitioning of the free energy of association between energetic and entropic contributions will also be critical in determining the temperature dependence of the association constants. Upon inspection of eqs. \eqref{eq:Lsflex}, \eqref{eq:Lflex}, and \eqref{eq:Linflex} in comparison to eq. \eqref{eq:LL}, we see that the entropy of association can take on the following forms depending on the flexibility of the associations. For semi-flexible associations, we have
\begin{align}
    \Delta S_{ij}=k_B\ln \left( \frac{(Z-2)^{2g}}{Z g^{2g} (1-g)^{2(1-g)}}\right).
\end{align}
For completely flexible associations, we have
\begin{align}
    \Delta S^{flex}_{ij}=k_B\ln \left( \frac{(Z-1)^{2}}{Z}\right).
\end{align}
Finally, for completely inflexible associations, we have
\begin{align}
    \Delta S^{inflex}_{ij}=-k_B\ln \left(Z\right)
\end{align}
Thus, for fully flexible, associations the entropy of association is strictly positive, and for fully inflexible associations the entropy of association is strictly negative. As we have explained, within the SCA, the entropy of association will always cancel out, and it will not affect the temperature dependence of $\tilde{\lambda}$ or the critical gel line. However, if the SCA is relaxed, the magnitude of the entropy of association will heavily affect  the temperature dependence of $\lambda_{ij}$, and thus will affect the critical gel line. 

In order to elucidate this point,  we can compute the gel line in the temperature-concentration plane for LiTFSI electrolyte. In the main text, $\tilde{\lambda}$ was found to be 0.075 for LiTFSI electrolyte at 300K. We  may also compute the values of the individual association constants using association probabilities computed from MD simulation (from eqs. \eqref{eq:sys3} and \eqref{eq:sys4}). For LiTFSI electrolyte at 300K, we found that $\lambda_{+-}=159$ and $\lambda_{+0}=2052$. Thus, as was explained in the previous section, these large association constants clearly indicate that the SCA will be appropriate for our system. However, when the temperature is increased both  $\lambda_{+-}$ and $\lambda_{+0}$ decrease depending on the magnitude of the entropy of association. 

\begin{figure*}[hbt!]
    \centering
    \includegraphics[width=\textwidth]{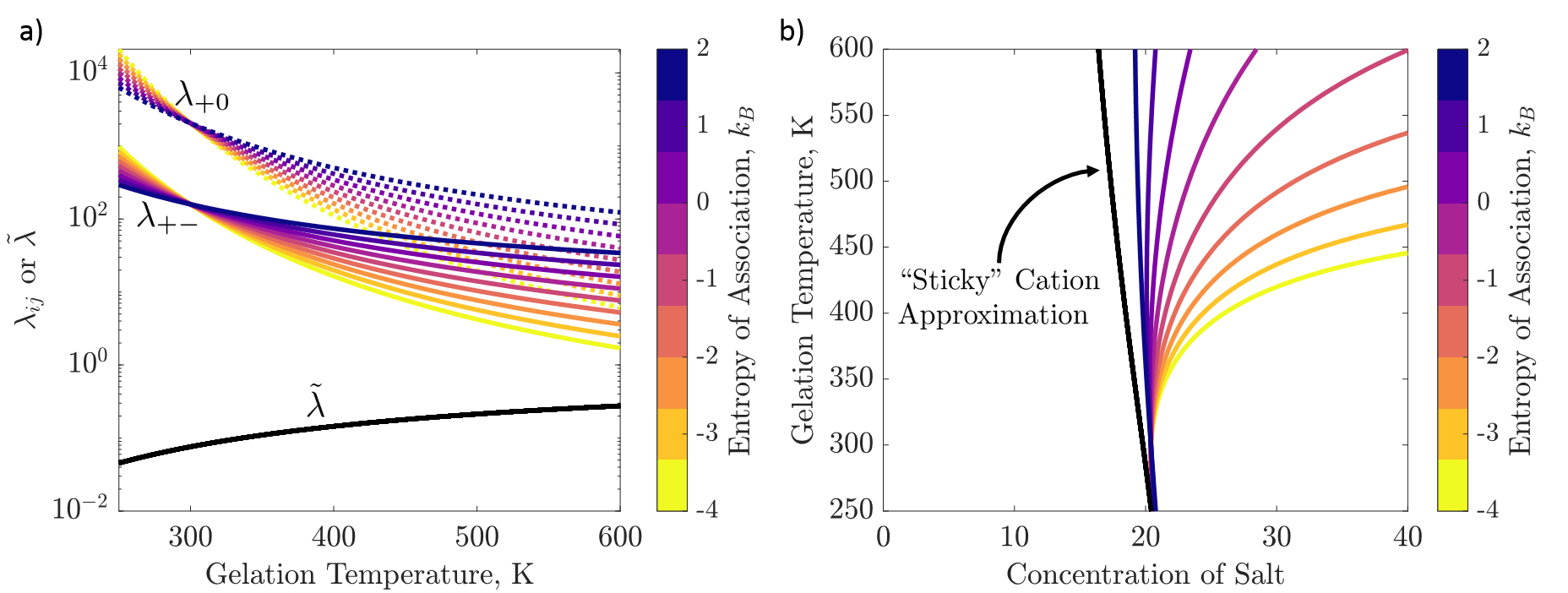}
    \caption{a) We plot the association constants for cation--anion association ($\lambda_{+-}$, solid, colored curves) and cation--solvent association ($\lambda_{+0}$, dotted, colored curves) as a function of temperature, for a range of different association entropies (2-4$k_B$). Additionally, we plot the reduced association constant ($\tilde{\lambda}$, black curve, used in the sticky cation approximation) as a function of temperature. b) We plot the critical gel line in the temperature--concentration plane, using the sticky cation approximation (black curve), and the general case (allowing for empty association sites on the cation) for a range of different association entropies (2-4$k_B$).}
    \label{fig:gel_boundary}
\end{figure*}

This can be seen in fig. \ref{fig:gel_boundary}a, where the $\lambda_{+-}$ (colored solid curves) and $\lambda_{+0}$ (colored dotted curves) are plotted as a function of temperature for various association entropies ranging from $-4k_B$--$2k_B$. Generally, it can be seen in fig. \ref{fig:gel_boundary}a that as the entropy of association decreases, the the association constants will decreases more drastically as a function of increasing temperature. For  reference, in ref. \cite{mceldrew2018}, the temperature dependent studies of clustering in emimBF$_4$ yielded an association entropy of -3.3$k_B$. A full temperature dependent study of clustering would be necessary to determine the entropy of association for the WiSEs studied here. On the contrary, $\tilde{\lambda}$ actually  increases as a function of temperature--in the high temperature limit $\tilde{\lambda}$ would tend toward 1. 

For increasing temperature, the SCA actually predicts increasing ionic association. Thus, the black curve in fig. \ref{fig:gel_boundary}b shows that the critical gel concentration mildly decreases as a function of increasing temperature.  This is a counter-intuitive result. Essentially, because the SCA conserves associations (e.g. every cation--anion association that is formed requires a cation--solvent association to be broken) the overall translational degrees of freedom of the mixture are conserved whether the electrolyte favors cation--anion or cation--solvent association. This means that the SCA predicts that there will be no entropic driving force to favor either type of association. Translational entropy is not increased by forming or breaking ion associations. Instead, for the SCA, increasing temperature only scales the energetic driving force for associations. Because $\Delta U_{+0}<\Delta U_{+-}$<0, the reduced cation--solvent association energy ($\Delta u_{+0}=\beta \Delta U_{+0}$) increases more rapidly as a function of temperature than the reduced cation--anion association energy ($\Delta u_{+-}=\beta \Delta U_{+-}$).

On the other hand, if we relax the SCA, we see that association (both cation-water and cation-anion association) decreases as a function of increasing temperature. This results in the critical gel concentration increasing with temperature, when the entropy of association is negative (expected for inflexible associations).

\subsection{Post Gel Association Probabilities}

Once the critical gel threshold has been surpassed, we must determine the volume fractions of cations, anions, and water in the gel, $\phi_+^{gel}$, $\phi_-^{gel}$, and $\phi_0^{gel}$ respectively. In order to do this, we employ Flory's treatment of the post-gel regime in which the volume fraction of free ions can be written equivalently in terms of overall association probabilities, $p_{ij}$, and association probabilities taking into account only the species residing in the sol, $p^{sol}_{ij}$

\begin{align}
    \phi_+(1-p_{+-})^{f_+}=\phi_+^{sol}(1-p^{sol}_{+-})^{f_+}
    \label{eq:gel1}
\end{align}
\begin{align}
    \phi_-(1-p_{-+})^{f_-}=\phi_-^{sol}(1-p^{sol}_{-+})^{f_-}
    \label{eq:gel2}
\end{align}
\begin{align}
    \phi_0(1-p_{0+})=\phi_0^{sol}(1-p^{sol}_{0+})
    \label{eq:gel3}
\end{align}
Where $\phi_i^{sol}=1-\phi_i^{gel}$ is the volume fraction of species $i$ in the sol. We may determine each of the three unknown sol volume fractions ($\phi_+^{sol}$, $\phi_-^{sol}$, $\phi_0^{sol}$), as well as the four unknown sol association probabilities ($p_{+-}^{sol}$, $p_{-+}^{sol}$, $p_{+0}^{sol}$, $p_{0+}^{sol}$) using Eqs.~\eqref{eq:gel1}-\eqref{eq:gel3} in addition to Eqs.~\eqref{eq:sys1}-\eqref{eq:sys2} \& Eqs.~\eqref{eq:sys3real}-\eqref{eq:sys4real}, using sol-specific quantities in this case.

Thus, we have 7 equations and 7 unknowns (four sol association probabilities and three sol species volume fractions). The fraction of species, $i$, in the gel is simply given by 
\begin{align}
    w_i^{gel}=1-\phi^{sol}_i/\phi_i
\end{align}
Note that prior to the critical gel concentration, we have the trivial solution that $p_{\pm\mp}=p^{sol}_{\pm\mp}$ and $\phi_i=\phi^{sol}_i$, yielding a gel fraction of $w_i^{gel}=0$. However, beyond the gel point, there is a non-trivial solution yielding $w_i^{gel}>0$.

\subsection{Theoretical Expressions for Activity}
Before deriving our model's expressions for activity it is useful to define the various related thermodynamic quantities. In general, the chemical potential of any species $i$ can be parsed as the following:
\begin{align}
    \mu_i&=\mu_i^{\theta}+k_BT \ln a_i \nonumber \\
    &=\mu_i^{\theta}+k_BT \ln\left(\frac{c_i}{c_i^{\theta}}\right)+\mu_i^{ex} \nonumber \\
    &=\mu_i^{\theta}+k_BT \ln \left(\frac{c_i}{c_i^{\theta}}\right)+k_BT \ln \gamma_i
\end{align}
where $\mu_i^{\theta}$ is the reference state chemical potential,  $a_i$ is the activity, $\mu_i^{ex}$ is the excess chemical potential, $\gamma_i$ is the activity coefficient, $c_i$ is the concentration (arbitrary units) and $c_i^{\theta}$ is the concentration in the reference state of species $i$. The activity is thus broken down as $a_i=\gamma_ic_i/c_i^{\theta}$, and the excess chemical potential has the definition $\mu_i^{ex}=k_BT\ln\gamma_i$. For ionic species this reference state chemical potential is taken to be at ``infinite dilution". Of course at infinite dilution we run into the problem that the concentration is zero, and thus the chemical potential would tend toward $-\infty$. In order to avoid this, it is standard to define the reference concentration to be 1 (arbitrary units). In our case, the reference chemical potentials are defined prior to mixing, and thus the reference concentration corresponds to species' volume fractions of 1. The chemical potential of a rank $lms$ cluster was written in Eq.~\eqref{eq:muclust}. 
As in Ref.~\citenum{mceldrew2020theory}, we may write the ionic activities using Eq.~\eqref{eq:muclust} and subtract off non-zero excess chemical potential 
\begin{align}
    \log a_+ =\beta (\mu_{100}- \mu^{\theta}_{100})= \beta \mu_{100}
     - \ln \left\{(1-p^\theta_{+-}-p^\theta_{+0})^{f_+}\right\}-(1-\xi_+ c^\theta_{tot})-\xi_+ \phi^{\theta}_0 d^{\theta}
\end{align}
for cations and
\begin{align}
    \log a_- =\beta (\mu_{010}- \mu^{\theta}_{010})= \beta \mu_{010}
     - \ln \left\{(1-p^\theta_{-+})^{f_-}\right\}-(1-\xi_- c^\theta_{tot})-\xi_- \phi^{\theta}_0 d^{\theta}
\end{align}
for anions where the $\theta$ again corresponds to the value at infinite dilution, $p^\theta_{-+}=p^\theta_{+-}=d^{\theta}=0$ and $\phi_0^{\theta}=c^{\theta}_{tot}=1$. Thus, we may write the ionic activities explictly as:
\begin{align}
    \ln a_+ =  \ln \phi_+ + f_+ \ln \left\{\frac{(1-p_{+-}-p_{+0})}{1-p^\theta_{+-}-p^\theta_{+0}}\right\} +\xi_+(1-c_{tot}+\phi_0 d)
    \label{eq:a+}
\end{align}
for cations and 
\begin{align}
    \ln a_- = \ln \phi_- + f_- \ln \left\{(1-p_{-+})\right\} +\xi_-(1-c_{tot}+\phi_0 d)
    \label{eq:a-}
\end{align}
for anions. In Eq.~\eqref{eq:a+}, we have the ratio of two divergent terms: $(1-p_{+-}-p_{+0})/(1-p^\theta_{+-}-p^\theta_{+0})$. This ratio is finite, however, and can be specified via Eq.~\eqref{eq:sys4} as the following:
\begin{align}
\frac{(1-p_{+-}-p_{+0})}{1-p^\theta_{+-}-p^\theta_{+0}}=\frac{p_{+0}}{1-p_{0+}} 
= \frac{1-p_{+-}}{1-p_{0+}}
\end{align}
Thus, the cation activity is written as
\begin{align}
    \ln a_+ = \ln \phi_+ + f_+ \ln \left\{\frac{1-p_{+-}}{1-p_{0+}}\right\} +\xi_+(1-c_{tot}+\phi_0 d)
    \label{eq:a+f}
\end{align}
The activity of water, $a_0$, is simply expressed as  
\begin{align}
    \ln a_0 =\beta \mu_{001}= \ln \phi_0 + \ln \left\{(1-p_{+0})\right\} +1-c_{tot}+\phi_\pm d.
    \label{eq:aw}
\end{align}
In Eq.~\eqref{eq:aw}, we do not need to adjust the activity to the infinite dilution reference state as with the ions, because the mixing reference state for water is already at the appropriate infinite dilution reference state. 
The final thing we must specify is $d$, which we recall has the definition: $d=\Delta^{gel\prime}_+c_+^{gel}+\Delta^{gel\prime}_-c_-^{gel}+\Delta^{gel\prime}_0c_0^{gel}$, where the $\prime$ notiation corresponds to a derivative with respect to $\phi_\pm$. Thus we need to know the functional dependence of  $\Delta^{gel}_i$. This is accomplished by enforcing an equilibrium between species in the sol and species in the gel ($\mu_i=\mu_i^{gel}$) and solving for  $\Delta_i^{gel}$ (again making the chemical potential reference state that at infinite dilution):
\begin{align}
    \Delta^{gel}_+(\phi_\pm)=\ln \phi_+ + f_+ \ln \left\{\frac{1-p_{+-}}{1-p_{0+}}\right\} +\xi_+
    \label{eq:dg1}
\end{align}
\begin{align}
    \Delta^{gel}_-(\phi_\pm)=\ln \phi_- + f_- \ln \left\{(1-p_{-+})\right\} +\xi_-
    \label{eq:dg2}
\end{align}
\begin{align}
    \Delta^{gel}_0(\phi_\pm)&=\ln \phi_0 + \ln \left\{(1-p_{0+})\right\} +1
    \label{eq:dg3}
\end{align}
Thus, we may simply plug in these expressions to $d$, obtaining the thermodynamic activities of each species used to generate the theoretical curves in Fig. 8 in the main text. 

\subsection{Additional Results for LiOTF and LiFSI}

In the main text, we discussed molecular simulation results and comparisons with the developed theory, but we primarily focused on LiTFSI electrolyte. Here we will outline additional results and comparisons for LiOTF and LiFSI electrolyte, as well. 
\begin{figure}[hbt!]
    \centering
    \includegraphics[width=\textwidth]{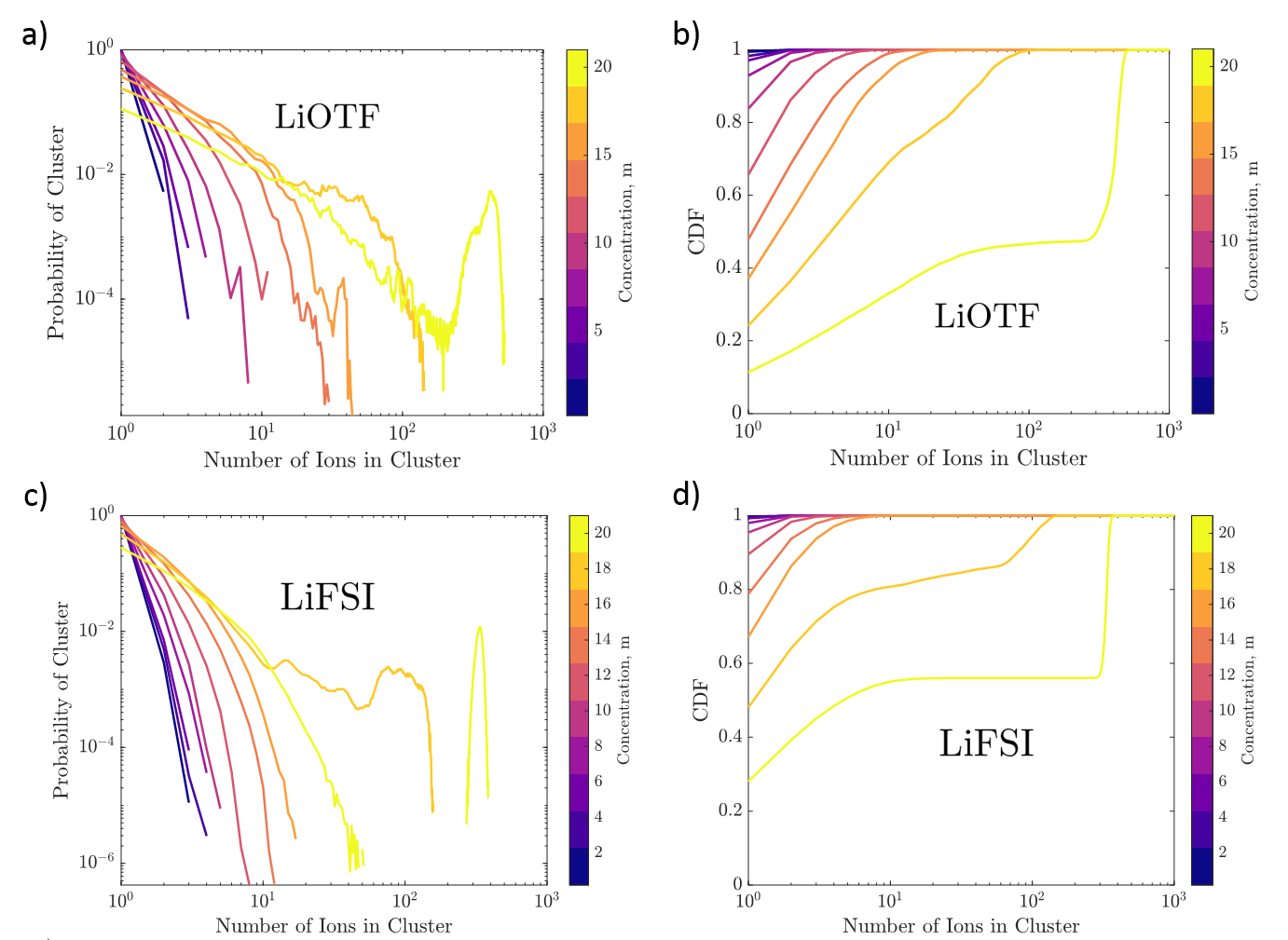}
    \caption{The cluster size probability distributions are plotted for a manifold salt concentrations for (a) LiOTF and (c) LiFSI  electrolytes. The cluster size cumulative distributions are also plotted for a manifold salt concentrations for (b) LiOTF and (d) LiFSI electrolytes.}
    \label{fig:agg_fsi_otf}
\end{figure}

In Fig.~\ref{fig:agg_fsi_otf}, we show the probability and cumulative distribution of clusters as a function of cluster size (total number of ions) for various salt concentrations for both LiOTF (Figs.~\ref{fig:agg_fsi_otf}a\&b) and LiFSI (Figs.~\ref{fig:agg_fsi_otf}c\&d). The general trends for the LiOTF and LiOTF electrolytes are very similar to the LiTFSI electrolyte discussed in the main text. In both LiFSI and LiTFSI electrolyte, there is a clear secondary peak in the cluster probability distribution indicating the presence of a percolating ion network when the salt concentration is 21~m. For LiFSI, there also is a subtle secondary peak (located at a cluster size  of roughly $10^2$ ions) for 15~m salt concentration. As we have mentioned in the main text, such a peak indicates the existance of a percolating cluster for 15~m LiFSI, which is not expected from the theory until about 21m, as LiFSI was found to be the least associating salt that we have studied. A possible explanation for the observation of percolation at such a low concentration in LiFSI could be the finite simulation box size, which would tend to induce percolation prematurely as compared to an infinite or macroscopic system. 
\begin{figure}[hbt!]
    \centering
    \includegraphics[width=\textwidth]{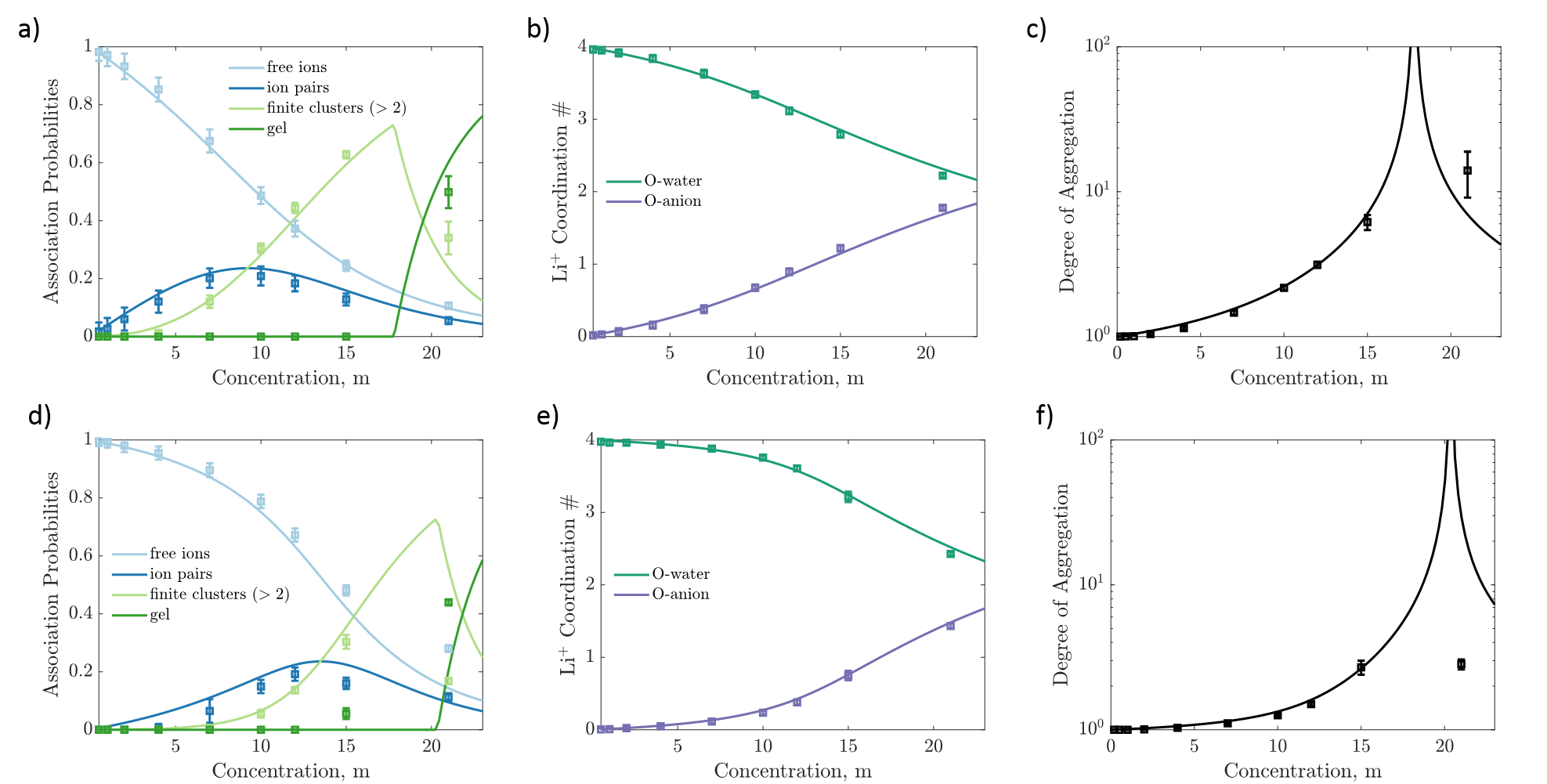}
    \caption{The probability distributions of cluster sizes (total number of ions in the cluster) are plotted for various salt concentrations in (a) LiOTF and (c) LiFSI electrolyte. The cumulative distributions of cluster sizes (total number of ions in the cluster) are plotted for various salt concentrations in (b) LiOTF and (d) LiFSI electrolyte. Theoretical parameters for LiOTF were $\lambda=0.184$, $\xi_+=0.4$, $\xi_-=5.9$, $f_+=4$, and $f_-=3$. Theoretical parameters for LiFSI were $\lambda=0.049$, $\xi_+=0.4$, $\xi_-=6.8$, $f_+=4$, and $f_-=3$}
    \label{fig:comp_fsi_otf}
\end{figure}

In Fig.~\ref{fig:comp_fsi_otf}, we show theory and MD simulation comparisons for LiOTF electrolyte (Fig.~\ref{fig:comp_fsi_otf}a-c) and for LiFSI electrolyte (Fig.~\ref{fig:comp_fsi_otf}d-f). Overall, the theory is able to quantitatively reproduce the ion cluster fractions (free ions, ion pairs, high order finite clusters, and gel), the Li$^+$ coordination numbers, and degree of aggregation as functions of salt concentration for both LiOTF and LiFSI electrolytes. Recall, our theoretical curves required the computation of the association constant $\tilde{\lambda}$ from simulations via the ionic association probabilities plotted in Fig.~4 in the main text. 
\subsection{Ion Association and Free Ion Lifetimes}
\begin{figure}[hbt!]
    \centering
    \includegraphics[width=\textwidth]{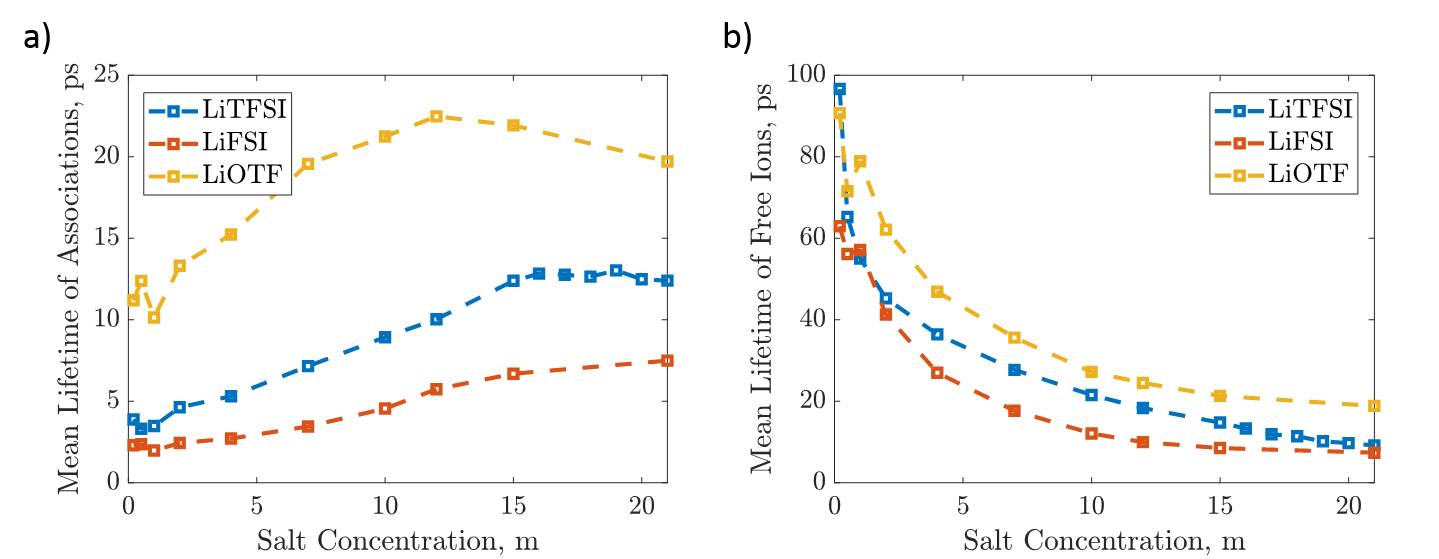}
    \caption{The mean lifetimes in picoseconds of (a) ion associations and (b) free ions are plotted as a function of salt concentration for aqueous LiTFSI (blue), LiFSI (red), and LiOTF (yellow)  electrolytes}
    \label{fig:lifetime}
\end{figure}

In Fig. \ref{fig:lifetime}, we plot the mean lifetimes of ion associations and free ions as a function of salt concentration for aqueous LiTFSI, LiFSI, and LiOTF electrolytes. 

\subsection{Molecular Dynamics Simulation Methodology}
In this study, we performed all-atom classical MD simulations using LAMMPS~\cite{plimpton1995}. We performed a set of simulations for three different salts: LiTFSI, LiFSI and LiOTF. For LiFSI and LiOTF salts we performed a series of fully periodic simulations at molal concentrations of 0.28~m, 0.5~m, 1~m, 2~m, 4~m, 7~m, 10~m, 12~m, 15~m, and 21~m. For LiTFSI salt we peformed the periodic simulations at molal concentrations of 0.28~m, 0.5~m, 1~m, 2~m, 4~m, 7~m, 10~m, 12~m, 15~m, 16~m, 17~m, 18~m, 19~m, 20~m, and 21~m.  

Each simulation contained 1000 water molecules and enough ion pairs to created as close to the specifies concentrations (0.2~m-21~m) as possible. The simulations were performed at fixed temperature (300~K) and pressure (1 bar), with Nose-Hoover thermostat and barostat until the density of the fluid equilibrated (10~ns with 1~fs time steps) and then, production runs were performed for 20~ns. The initial configurations for all simulations were generated using the open-source software, PACKMOL~\cite{martinez2009packmol}. All MD simulations were visualized using the open-source software, VMD~\cite{HUMP96}.

For all ionic species we employed the CL$\&$P force field, which was developed for ionic liquid simulations, with same functional form as the OPLSAA force field~\cite{lopes2012}. Given the dense ionic nature of our systems, we expect the CL$\&$P force field to be appropriate for WiSEs. For water, we employed the spc/e force field. Interatomic interactions are determined using Lorentz-Berthelot mixing rules. Long range electrostatic interactions were computed using the Particle-Particle Particle-Mesh (PPPM) solver (with a cut-off length of 12~$\AA$). 

\subsection{Computing Activity from Molecular Dynamics}
\begin{figure}[hbt!]
    \centering
    \includegraphics[width=0.6\textwidth]{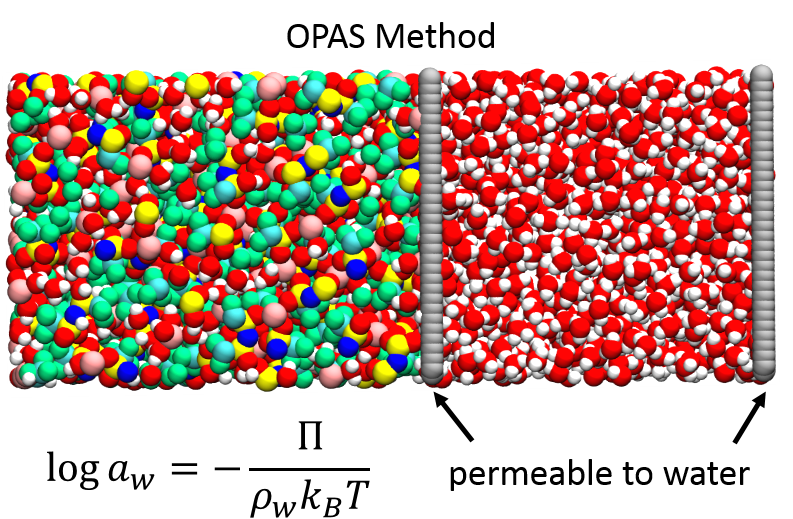}
    \caption{A schematic of the MD simulation box set-up used to employ the OPAS method for computing solvent activity. The dimensions of the chamber on the right containing solely water are 30~$\AA$ $\times$ 30~$\AA$ $\times$ 33.2~$\AA$. The box on the left containing the electrolyte has the dimensions 30~$\AA$ $\times$ 30~$\AA$ $\times$ 41.7~$\AA$.}
    \label{fig:opas}
\end{figure}

In order to measure the activity of the species in our MD simulations, we employed the OPAS (osmotic pressure for activity of solvents) method developed in Ref.~\citenum{kohns2016solvent}, which involves measuring the osmotic pressure difference between a chamber containing only solvent molecules, and a chamber of an electrolyte of interest (see Fig.~\ref{fig:opas}). The electrolyte is described by the same force field as in the previously discussed fully periodic simulations. However, the chambers are separated by an FCC sheet of hard spheres, which prevent salt molecules from permeating, but allow solvent molecules to pass freely in between the chambers. The chamber dimensions are given in the caption of Fig.~\ref{fig:opas}. The system pictured in Fig.~\ref{fig:opas}, is fully periodic, but the dynamics is performed in an NVT ensemble, with a Nose-Hoover thermostat at a target temperature of 300~K. The system is simulated until the internal pressure and concentration distributions have equilibrated, and we designated 10~ns for this equilibration for each simulation. We then perform production runs of 10~ns for each simulated concentration in which the average pressure in each chamber is calculated. The initial number of water molecules in the water-only chamber is 1000 for all simulations. The number of ions and water molecules in the electrolyte chamber are given in Tbl.~\ref{tbl:5opas} for each simulated concentration. 
\begin{table}[!hbt]
\centering
\begin{tabular}{ll|ll|ll|}
\hline
Conc. (m) & $\quad$ & $\#$ of ion pairs & $\quad$ & $\#$ of waters  \\
\hline
0.5m && 9 && 1187  \\
1m && 18 && 1117  \\
2m && 36 && 1000  \\
4m && 72 && 822  \\
7m && 81 && 642  \\
10m && 97 && 539  \\
12m && 105 && 485  \\
15m && 113 && 418  \\
21m && 125 && 330  \\
\hline
\end{tabular}
\caption{The number of ion pairs and water molecules initially placed in the electrolyte chamber for the OPAS method simulations.}
\label{tbl:5opas}
\end{table}

The difference in average pressure yields the osmotic pressure $\Pi$, which can be related directly to the activity of the water, $a_0$ in the electrolyte chamber (using the Gibbs-Duhem Relation for the pure water chamber and an assumption of chemical equilibrium between the two chambers):
\begin{align}
    \log(a_0)=-\frac{\Pi}{\rho_0  k_BT}
\end{align}
where $\rho_0$ is the molar concentration of water in the chamber containing only water.

Upon computing the activity of the water in the electrolyte as a function of salt concentration, we can use the Gibbs-Duhem relation for the electrolyte chamber to compute chemical potential of the salt (with respect to a dilute reference state):
\begin{align}
    \Delta \mu_s(x) = -\frac{1}{2}\int\frac{1-x}{x}d\mu_0(x)
    \label{eq:gd}
\end{align}
where $x$ is the mole fraction of salt, and $\mu_s=(\mu_++\mu_-)/2$ is the mean chemical potential of the salt. $\Delta \mu_s(x) $ is the difference mean chemical potential of the salt from a reference salt concentration of 0.5m and that at a mole fraction of $x$. We can numerically compute the integral in Eq.~\eqref{eq:gd} using the simulated values of the activity of water [$d\mu_0=d\ln(a_w)$], however our numerical accuracy is greatly limited by the discreteness of the activity measurements. For this reason, we instead fit a polynomial (order 4) to the simulated activity and compute the integral exactly, yielding  the ``MD Gibbs-Duhem" curve in Fig.~8 in the Main text.

\bibliography{main.bib}

\end{document}